\documentclass[amsmath,amssymb,aps,pra,twocolumn,superscriptaddress,nofootinbib,floatfix]{revtex4-2}

\usepackage{graphicx}
\usepackage{subcaption}
\usepackage{dcolumn}
\usepackage{bm}
\usepackage{siunitx}
\usepackage[colorlinks=true,linkcolor=black,citecolor=blue,urlcolor=black]{hyperref}
\usepackage{enumitem}
\usepackage{comment}
\usepackage{mathtools}
\usepackage{bbold}
\usepackage{ragged2e}
\usepackage{float} 

\DeclarePairedDelimiter\norm{\lvert}{\rvert}   

\DeclareSIUnit\atomicmassunit{u}
\DeclareSIUnit\bar{bar}

\usepackage{color}
\usepackage{xcolor}

\begin{document}



\title{Deterministic engineering of topological defects in two-dimensional single-species ion Coulomb crystals}



\author{L.-A. Rüffert}
\affiliation{Physikalisch-Technische Bundesanstalt, Bundesallee 100, Braunschweig, 38116, Germany}

\author{H. Kalathur}
\affiliation{Physikalisch-Technische Bundesanstalt, Bundesallee 100, Braunschweig, 38116, Germany}

\author{T. E. Mehlstäubler}
\affiliation{Physikalisch-Technische Bundesanstalt, Bundesallee 100, Braunschweig, 38116, Germany}
\affiliation{Institut für Quantenoptik, Leibniz Universität Hannover, Welfengarten 1, Hannover, 30167, Germany}
\affiliation{Laboratorium für Nano- und Quantenengineering, Leibniz Universität Hannover, Schneiderberg 39, Hannover, 30167, Germany}

\date{\today}

\begin{abstract}
We propose and numerically demonstrate a deterministic protocol for preparing and tailoring topological defects in two-dimensional single-species Coulomb crystals confined in radio-frequency traps. In conventional finite-rate quenches through the linear-to-zigzag transition, defects form stochastically from independently chosen broken-symmetry domains, and the quench rate controls the resulting domain structure. We instead use an initial three-dimensional helix structure and planarize it into a deterministic defected state. We demonstrate that the number and approximate axial positions of kinks are encoded by the projected nodes of the initial helix.
We find that selected helices can remain on metastable finite-winding branches during planarization and yield reproducible single- or multi-kink states, even in the adiabatic limit. Using the quench rate as an additional control parameter substantially enlarges the set of precursor helices that result in defected states, as a fast planarization can prevent reordering during the quench and freeze in the pre-existing domain pattern.
After planarization, the survival and mutual interaction of the defects are governed by their effective Peierls--Nabarro potential landscape. By numerically isolating the interaction energy for the two defect types, we find that the interaction of extended kinks is attractive over the investigated parameter range, while odd kinks exhibit a short-range repulsive barrier tunable by the anisotropy of the trapping potential $\alpha$. 
Finally, we introduce an experimentally motivated post-selection scheme in which $\alpha$ is used to selectively destabilize, annihilate, or convert targeted defects. Our results provide a route to deterministic single- and multi-kink preparation in single-species Coulomb crystals, enabling controlled studies of defect interactions, transport, and nanofriction.
\end{abstract}

\maketitle

\section{introduction}
\label{sec:Introduction}
Laser-cooled trapped ions confined in radio-frequency (rf) Paul traps self-organize into ordered Coulomb crystals once the thermal energy becomes small compared to the mutual Coulomb repulsion \cite{Drewsen_IonCoulombCrystals_2015a, drewsen_LargeIonCrystals_1998, raizen_IonicCrystalsLinear_1992}. 
These trapped-ion crystals constitute a versatile platform spanning precision metrology and optical clocks \cite{ludlow_OpticalAtomicClocks_2015, rosenband_FrequencyRatioHg_2008, schmidt_Spectroscopy_2005a, brewer_QuantumLogicClock_2019}, quantum information processing \cite{ludlow_OpticalAtomicClocks_2015, cirac_QuantumComputations_1995a, wineland_ExperimentalIssuesCoherent_1998, kielpinski_LargescaleIontrap_2002, bruzewicz_TrappedionQuantumComputing_2019}, and quantum simulation of many-body dynamics \cite{blatt_QSimulationsTrapped_2012, schneider_QuantumSimulations_2012, porras_EffectiveQuantumSpin_2004, friedenauer_SimulatingQuantumMagnet_2008, britton_EngineeredTwodimensionalIsing_2012a}. 
In addition, ion Coulomb crystals provide a realization of strongly correlated mesoscopic matter that enables controlled studies of structural transitions and non-equilibrium dynamics in one, two, and three dimensions \cite{morigi_IonCoulombCrystals_2026, dubin_TrappedNonneutralPlasmas_1999a, Schiffer_PhaseTransitions_1993, ruffert_DomainFormationStructural_2024c, Fishman_Phase_Transitions_2008}.

An important aspect of these systems is the existence of topological defects that can emerge when a crystal is driven through a symmetry-breaking transition. In linear Paul traps, reducing the radial confinement causes a transition from a one-dimensional (1D) linear chain to a planar two-row zigzag structure. When this transition is crossed non-adiabatically, different zigzag domains may form with incompatible transverse orientation, creating localized topological defects (``kink'' solitons) at the domain boundaries \cite{delcampoStructural2010, PykaDefectFormation2013, ulm_ObservationKibbleZurek_2013, PartnerDefects2013a, landa_StructureDynamicsBifurcations_2013, mielenz_TrappingDefects_2013}.
 
These kinks are nonlinear excitations with spectrally separated modes, allowing for selective addressing and long-lived coherent dynamics~\cite{mielenz_TrappingDefects_2013,landaQuantumCoherenceDiscrete2010,brox_SpectroscopyTransport_2017}. Trapped ion chains hosting topological defects enable controlled investigations of nanofriction and symmetry-breaking transitions in finite systems, including the Aubry-type transition~\cite{bylinskii_AubryTransition_2016, kiethe_Aubrytype_2017a}.
Moreover, kinks provide an attractive platform for controlled studies of non-equilibrium dynamics, ranging from defect-mediated energy transport~\cite{timm_EnergyLocalization_2020} to genuinely quantum effects such as entanglement of localized motion~\cite{landa_Entanglement_2014} and quantum nanofriction~\cite{timm_QuantumNanofriction_2021a}.

In this work, we focus exclusively on defects in planar 2D zigzag crystals formed in linear rf traps. However, topological defects are not restricted to planar zigzag crystals but can also appear in more complex 2D and 3D ion configurations \cite{radzvilavicius_TopologicalDefect_2011a, dubin_TwodimensionalPlasma_2013, mielenz_TrappingDefects_2013, Nigmatullin_Helical_2016, arnold_DynamicsVortexDefect_2022, morigi_IonCoulombCrystals_2026}. 

In conventional linear-to-zigzag quenches, kinks arise from incompatible local choices of the broken-symmetry zigzag orientation with their number and positions being stochastic quantities governed by Kibble--Zurek-type dynamics~\cite{delcampoStructural2010,ulm_ObservationKibbleZurek_2013,PykaDefectFormation2013}.
Stable two-kink configurations have nevertheless been observed in sufficiently large single-species crystals, where kink–kink interactions can suppress annihilation~\cite{landa_StructureDynamicsBifurcations_2013}.
Additional pinning mechanisms, such as mass defects, provide another route to stabilizing multiple kinks at distinct positions~\cite{delcampoStructural2010,ulm_ObservationKibbleZurek_2013,PykaDefectFormation2013}. 

Here, we use molecular dynamics simulations to demonstrate a deterministic and controlled route to prepare single and multiple kinks in single-species Coulomb crystals. Starting from a prepared three-dimensional (3D) helix structure, a planarization quench maps the crystal into a planar zigzag configuration and kinks can nucleate at the positions of the projected helical nodes. The resulting defect number and placement can be tuned by the geometry of the initial helix and the chosen quench basis. Selected helices can remain on metastable finite-winding branches during planarization and yield reproducible single- or multi-kink states, even in the adiabatic limit. When the quench rate is
used as a control parameter, the set of precursor helices that result in a defected state is substantially enlarged, as the fast planarization can reliably freeze in the pre-existing domain pattern of the helix into the planar phase.

In addition, we investigate kink--antikink interaction for different defect types and isolate an effective interaction energy from the two-kink Peierls--Nabarro potential, which describes the potential landscape of the system with respect to the collective kink coordinates. We find that while the interaction of extended kinks is attractive over the investigated parameter range, odd kinks exhibit a distinct repulsive barrier which is tunable by the aspect ratio of the trapping potential $\alpha$. This makes the kink--antikink interaction a tunable variable in multi-defect experiments.

Lastly we provide an experimentally motivated post-selection method for multi-kink crystals, by showing how defects can be selectively destabilized and annihilated using $\alpha$ as a control parameter. This approach can be used to yield a desired kink configuration from an initial multi-kink state as well as transforming odd kinks into extended kinks and vice versa.

Our results provide a practical strategy to prepare reproducible multi-kink states in finite, impurity-free Coulomb crystals, tune their mutual interaction and create desired multi-defect configurations using targeted destabilization and conversion between kink-types. These findings are relevant for controlled studies of interacting topological defects in self-organized Coulomb crystals, including defect-mediated energy transport, engineered multi-kink arrangements and nanofriction phenomena.

\section{theoretical background}
\label{sec:Theory}

\subsection{Ion Coulomb crystals}
\label{subsec:ion_coulomb_crystals}
When ions are confined in a Paul trap and laser-cooled to temperatures of a few millikelvin, they arrange into ordered structures known as Coulomb crystals~\cite{raizen_IonicCrystalsLinear_1992, drewsen_LargeIonCrystals_1998, dubin_TrappedNonneutralPlasmas_1999a, Drewsen_IonCoulombCrystals_2015a}.
Confinement in a Paul trap is achieved by applying a fast oscillating radio-frequency (RF) electric field. The net effect on the ions can be described in the ponderomotive approximation as a static harmonic potential in all three spatial directions~\cite{paul_1990}.

In the following, we study a system of $N$ identical ions with mass $m$ and charge $Q$, whose positions are denoted by $\vec r_i=(x_i,y_i,z_i)$.
Throughout this work, we use a fixed laboratory coordinate system in which the $z$ axis defines the axial direction of the trap, while $x$ and $y$ span the radial plane. The $y$ axis is aligned with the detection line of sight, so that the observable imaging plane is the $x$-$z$ plane.

The ions interact through the Coulomb force, such that the total potential energy is
\begin{align}\label{eq:potential}
\mathcal{V}
&= \sum_{i=1}^N \frac{m}{2}\Bigl(\omega_x^2 x_i^2 + \omega_y^2 y_i^2 + \omega_z^2 z_i^2\Bigr)
+ \sum_{i<j}^N \frac{Q^2}{4\pi\epsilon_0\, d_{ij}}\,,
\end{align}
where $\epsilon_0$ is the vacuum permittivity and $d_{ij}=\norm{\vec r_i-\vec r_j}$ denotes the separation of ions $i$ and $j$.
The secular frequencies $\omega_x$, $\omega_y$, and $\omega_z$ quantify the strength of confinement along the three axes and therefore control the crystal’s aspect ratio. They are given by
\begin{equation}
\begin{aligned}
\omega_{z,i}^2 &= \frac{Q_i}{m_i}\,u_\text{DC}, \\
\frac{\omega_{x/y}^2}{\omega_{z}^2} &= \frac{1}{2}\frac{Q}{m}\frac{u_\text{RF}^2}{u_\text{DC}\,\Omega_\text{RF}^2}
- \frac{1}{2} \mp c_{xy}\,,
\end{aligned}
\label{eq:omegas}
\end{equation}
with $u_\text{DC}$ and $u_\text{RF}$ denoting the gradients of the static and RF electric fields, $\Omega_\text{RF}$ the RF drive frequency, and $c_{xy}$ accounting for residual anisotropies within the radial plane.

In the limit of vanishing temperature, the ions minimize $\mathcal{V}$ and occupy equilibrium positions $\{\vec r_i^{\,0}\}$ determined by the competition between mutual Coulomb repulsion and the trap potential. 
Throughout this work we focus on the regime
\begin{equation}
\omega_z \ll \omega_x \le \omega_y\,,
\end{equation}
which will restrict planar crystal structures into the $x$-$z$-plane.
It is useful to define the aspect ratio of the trapping frequencies as
\begin{equation}
    \alpha=\frac{\omega_x}{\omega_z}\, ,
\end{equation}
which acts as a control parameter to drive the structural phase transitions.

To make our results independent of a particular ion species or a specific experimental trap setting, we express positions and energies in dimensionless units. This is done by normalizing energies with
\begin{equation}
\varepsilon = m\,\omega_z^2\,\ell^2,
\label{eq:energy_scale}
\end{equation}
where $M$ denotes the ion mass and the associated characteristic length of the Coulomb crystal is chosen as
\begin{equation}
\ell=\left(\frac{Q^2}{4\pi\varepsilon_0\,m\,\omega_z^2}\right)^{1/3}.
\end{equation}
With these definitions, the results are independent of the chosen ion species and axial trap frequency and allow for comparisons between different experimental setups.

\subsection{Structural equilibrium phases and order-parameter manifolds}
\label{subsec:phase_transitions_order_param}
\begin{figure*}
  \includegraphics[width=\linewidth]{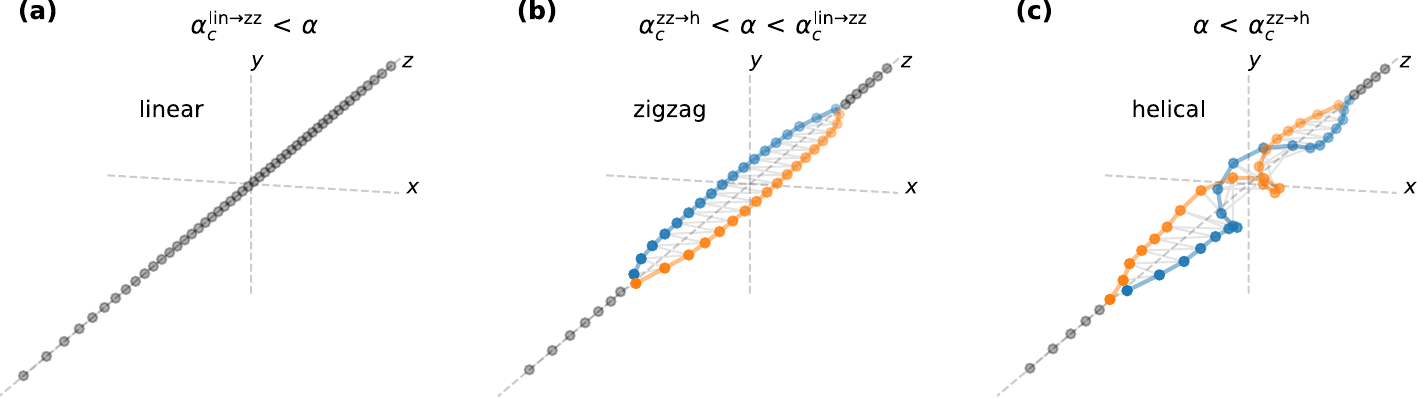}
  \caption{\justifying Structural configurations of a finite $N=50$ ion Coulomb crystal at different aspect ratios of the trapping frequencies $\alpha=\omega_x/\omega_z$ with $\omega_x=\omega_y$. \textbf{(a)} For $\alpha>\alpha_c^{\mathrm{lin}\to\mathrm{zz}}$, the crystal remains stable in a linear configuration. \textbf{(b)} For $\alpha_c^{\mathrm{zz}\to\mathrm{h}}<\alpha<\alpha_c^{\mathrm{lin}\to\mathrm{zz}}$, the ions buckle radially and form a zigzag pattern along $z$. Ions with a radial offset larger than $\sim1\%$ of the average ion spacing are colored alternatingly in blue and orange along $z$. \textbf{(c)} For $\alpha<\alpha_c^{\mathrm{zz}\to\mathrm{h}}$, the central zigzag region twists into a helical configuration, starting from the region of largest ion density near the trap center. The helical twist increases with decreasing $\alpha$.}
  \label{fig:phase_transition_schematic}
\end{figure*}
Structural transitions in ion Coulomb crystals are governed by the competition between the external trapping potential and the mutual Coulomb repulsion. For sufficiently strong radial confinement, the ions minimize their potential energy by forming a linear chain along the more weakly confined axial direction. When the radial confinement is reduced below a critical value, the transverse zigzag mode softens and destabilizes the linear chain. This linear-to-zigzag instability is a continuous structural phase transition of second order, with the order parameter being the transverse displacement from the $z$ axis~\cite{Fishman_Phase_Transitions_2008, Partner_Phase_Transitions_2015a}.

When considering the equilibrium structure of an ideal homogeneous Coulomb crystal in the planar case, where one radial direction is much more strongly confined than the other, e.g. $\omega_y \gg \omega_x$, the ions are restricted to the $x$-$z$ plane. For an ideal infinite planar zigzag, the equilibrium positions can be written as
\begin{equation}
    z_k^{(0)} = k a ,\qquad
    x_k^{(0)} = \sigma (-1)^k \phi_0 , \qquad
    \sigma=\pm 1 .
\end{equation}
Here, $a$ is the axial lattice spacing, $k$ denotes the ion index, $\phi_0$ is the zigzag amplitude, and $\sigma$ labels the two degenerate zigzag ground states ~\cite{Fishman_Phase_Transitions_2008}.

Equivalently, a macroscopic staggered order parameter can be defined as
\begin{equation}
    \Phi = \lim_{N\rightarrow\infty} \frac{1}{N} \sum_{k=1}^{N} (-1)^k x_k \, ,
    \label{eq:macro_order_param}
\end{equation}
providing a mean value to describe the phase of the system which is either 
\begin{equation}
    \Phi = 0
\end{equation}
in the linear phase or 
\begin{equation}
    \Phi = \sigma \phi_0 = \pm \phi_0 ,
\end{equation}
for the two homogeneous planar zigzag ground states. The planar linear-to-zigzag transition therefore breaks a discrete reflection symmetry, and the broken-symmetry manifold is $Z_2$.

In the radially symmetric case ($\omega_x=\omega_y$) the order parameter becomes complex and can be written as
\begin{equation}
    \Phi = \lim_{N\rightarrow\infty}\frac{1}{N}\sum_{k=1}^{N} (-1)^k (x_k+i y_k)=|\Phi| e^{i\Theta}.
    \label{eq:complex_order_parameter}
\end{equation}
Here, $|\Phi|$ is the zigzag amplitude and $\Theta$ describes the angle between the zigzag plane and the $x$ axis. In the broken-symmetry phase, all orientations $\Theta\in[0,2\pi)$ are degenerate. The broken-symmetry manifold is therefore the continuous circle $S^1$.

Therefore, the degeneracy of the broken-symmetry phase depends on the radial confinement, being $Z_2$ for a planar system ($\omega_x \ne \omega_y$) and $S^1$ in the radially symmetric case ($\omega_x = \omega_y$). Figure~\ref{fig:phase_transition_schematic} (a) and (b) show an example of the linear-to-zigzag transition for a finite Coulomb crystal of 50 $^{172}$Yb$^+$ ions in the radially symmetric case.

When further reducing the transverse confinement, the 2D zigzag structure undergoes a transition into a helical phase, illustrated in Fig.~\ref{fig:phase_transition_schematic} (c), followed by transitions to more complex 3D spheroidal crystal configurations~\cite{dubin_StructuralPhase_1993,Schiffer_PhaseTransitions_1993,piacente_Generic_2004,Fishman_Phase_Transitions_2008,landa_StructureDynamicsBifurcations_2013}. 

In finite harmonically confined ion crystals, the axial ion density is position dependent and largest near the trap center. Consequently, the local critical density is reached first in the central region, such that the zigzag or helical instability appears first at the center and then extends outward as the radial confinement is further reduced~\cite{Fishman_Phase_Transitions_2008}. The finite crystal therefore approaches the idealized phase sequence through spatially inhomogeneous configurations, with linear outer regions, a central zigzag domain, and, at lower radial confinement, a central helix.

\subsection{Local order-parameter fields and non-equilibrium kink formation}

Driving the continuous linear-to-zigzag transition at a finite rate can produce spatially varying order-parameter configurations which can be described by a local staggered transverse displacement field
\begin{equation}
    \phi_k = (-1)^k (x_k+i y_k) \, ,
    \label{eq:local_staggered_field}
\end{equation}
which is the discrete version of the complex Ginzburg--Landau order parameter \eqref{eq:complex_order_parameter} used for the linear-to-zigzag transition~\cite{Fishman_Phase_Transitions_2008, chiara_SpontaneousNucleation_2010, Nigmatullin_Helical_2016}. In the linear phase, $\phi_k=0$. After crossing the transition, $\phi_k$ becomes finite and locally selects an element of the broken-symmetry manifold.

In the planar case, where one radial direction is strongly confined ($\omega_x \ll \omega_y$), the local $Z_2$ order parameter reduces to the real staggered displacement
\begin{equation}
    s_k = (-1)^k x_k .
\end{equation}
For a homogeneous planar zigzag domain,
\begin{equation}
x_k = \sigma_k (-1)^k \phi_0 ,
\qquad
\sigma_k=\pm1 ,
\end{equation}
and therefore
\begin{equation}
s_k = \sigma_k \, \phi_0 \, .
\end{equation}
Within the same zigzag domain, $\sigma_k$ has the same sign. During finite-rate quenches over the linear-to-zigzag instability, spatial regions that are separated by more than the relevant correlation length fall out of equilibrium independently and $\sigma_k$ is chosen stochastically. Kinks form where incompatible domains meet, and the resulting kink density follows Kibble--Zurek-type scaling with the quench rate~\cite{delcampoStructural2010,PykaDefectFormation2013,ulm_ObservationKibbleZurek_2013,PartnerDefects2013a}. Depending on the aspect ratio of the trapping potential, planar defects can generally occur as odd or extended kinks, as detailed in Appendix~\ref{app:defect_regimes}.

A local topological charge can be assigned directly to the bond between neighboring domains as
\begin{equation}
q_k =
\frac{\sigma_{k+1}-\sigma_k}{2} \,
\in \,{-1,0,+1} , .
\end{equation}
A nonzero $q_k$ marks a domain wall, with the sign distinguishing kink and antikink depending on the chosen convention. 
Since neighboring domain walls of a $Z_2$ order-parameter field in one dimension necessarily carry alternating topological charges, adjacent kink--antikink pairs can annihilate when they approach each other. 
The interaction dynamics between odd and extended kinks are analyzed in more detail in Sec.~\ref{sec:kink_interaction}.

In finite harmonically confined crystals, additional processes modify the ideal Kibble--Zurek scaling. Kinks can be lost at the boundaries of the zigzag region or annihilate with other defects before the crystal reaches a steady state~\cite{PartnerDefects2013a}. Experiments using single-species crystals typically observe at most one long-lived defect per quench in the relevant parameter regime~\cite{PykaDefectFormation2013}. More complex multi-defect configurations can be stabilized by additional pinning mechanisms. In particular, mass defects can create local minima in the potential landscape that trap kinks, allowing multiple kinks to be stabilized in the same finite crystal~\cite{landa_StructureDynamicsBifurcations_2013, PartnerDefects2013a}. However, repeatable stable trapping of multiple defects in a finite single-species Coulomb crystal has yet to be demonstrated.

\subsection{Deterministic defect engineering by order-parameter projection}
\label{subsec:defect_engineering}
\begin{figure*}
  \includegraphics[width=\linewidth]{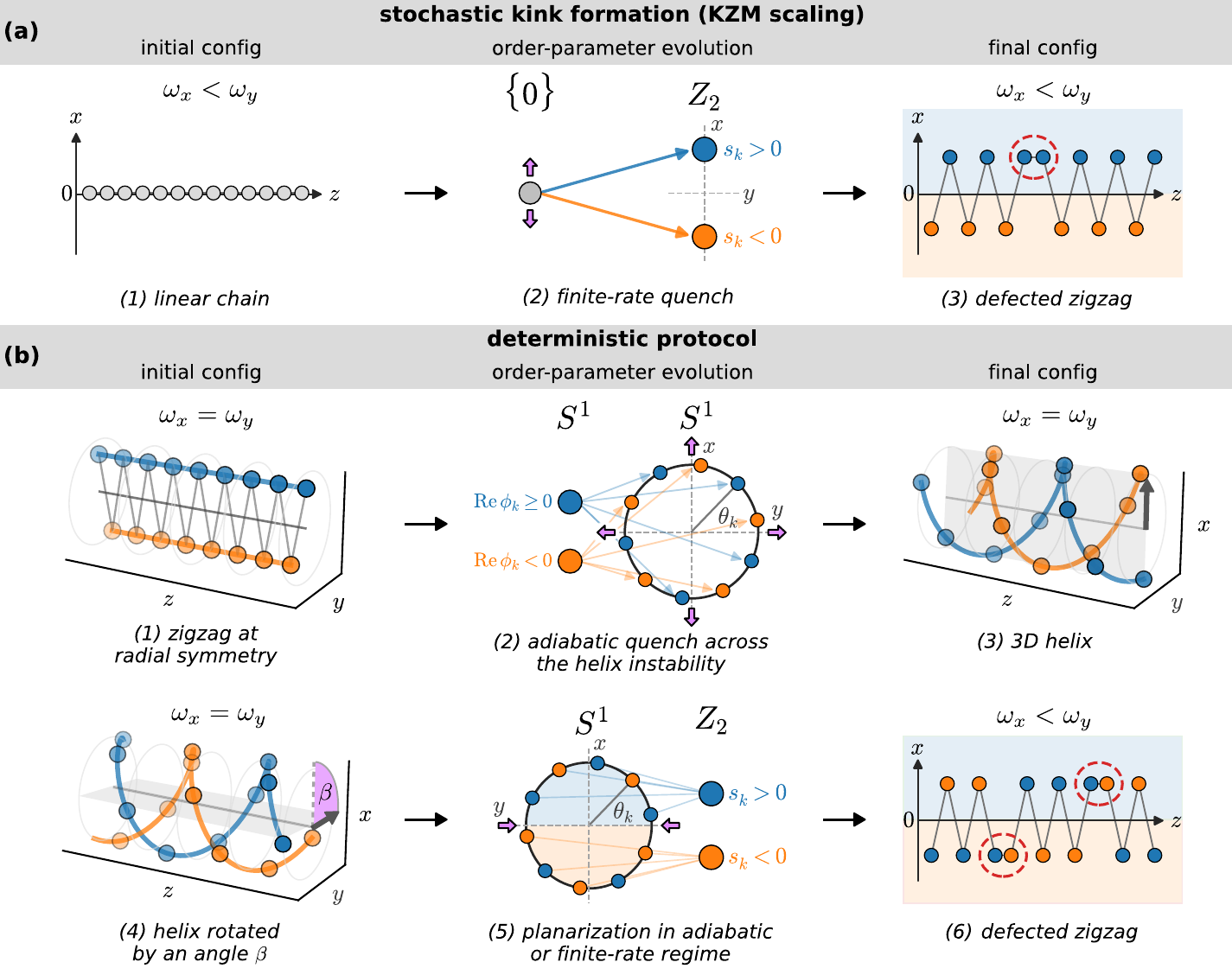}
  \caption{\justifying
Schematic comparison of stochastic kink formation and deterministic defect engineering.
\textbf{(a)} In a conventional finite-rate linear-to-zigzag quench, the initial linear chain (1) corresponds to the symmetric order-parameter state $\{0\}$. 
During the transition (2), different regions stochastically select one of the two planar $Z_2$ branches, indicated by $s_k>0$ and $s_k<0$. The purple arrows indicate the quench direction.
Incompatible choices lead to defects in the final zigzag configuration (3).
\textbf{(b)} In the deterministic protocol, a defect-free zigzag in a radially symmetric potential (1) is converted into a helix by adiabatically reducing the radial confinement across the helix instability (2) and thereby forming a three-dimensional helix (3). The gray shaded area indicates an $x$-$z$ reference plane of the initial helix. The helix is then rotated (4) by an angle $\beta$ relative to this reference plane.
Planarization along $y$ projects the complex order parameter from the $S^1$ manifold back into the real planar $Z_2$ manifold (5). This planarization quench is performed either in the adiabatic limit or with a finite quench rate. In both cases, the final state depends solely on the initial projection of the helix onto the chosen quench basis. The projected helix nodes are imprinted as deterministic domain walls and result in a single- or multi-defected zigzag (6).}
  \label{fig:deterministic_kink_formation}
\end{figure*}
The defect-formation mechanism discussed above relies on non-equilibrium symmetry breaking. During a finite-rate quench through the linear-to-zigzag transition, different domains can choose the broken-symmetry state independently, such that defects arise stochastically from incompatible local choices. 
Our approach follows a different route: The domain pattern is predetermined by encoding it in a three-dimensional helix. The precursor helix is then rotated by a controlled angle $\beta$ about the trap axis
relative to the $x$-$z$ detection plane and subsequently
planarized by increasing the confinement along the fixed $y$
direction. The quench outcome does not rely on stochastic
symmetry breaking and, for a fixed quench rate, depends only on the
initial helix configuration and its orientation $\beta$. This approach allows for deterministic preparation of multiple kinks in single-species Coulomb crystals, even in the adiabatic limit.
Figure~\ref{fig:deterministic_kink_formation} schematically contrasts this deterministic projection protocol with conventional stochastic kink formation by a finite-rate linear-to-zigzag quench.

As described in Sec.~\ref{subsec:phase_transitions_order_param}, a defect-free crystal in the planar zigzag phase can be driven into a helix state by lowering the radial confinement in the radially symmetric case, $\omega_x=\omega_y$. 
Although the idealized preparation assumes radial symmetry, exact degeneracy of the radial trap frequencies is not required to generate a three-dimensional helical precursor. Extended simulations show that crystal sizes on the order of $N\sim50$, a radial anisotropy of a few percent can still yield nearly uniform helices, while the exact tolerance depends on ion number and confinement. Larger anisotropies increasingly pin and deform the helix and can lead to nonuniform twisting with reduced or vanishing net winding. Such deformations do not by themselves preclude deterministic kink preparation proposed here, but strong anisotropy could limit the reproducibility of the initial configuration.

In the radially symmetric case, once the zigzag becomes unstable by lowering the radial confinement, the two zigzag chains reduce the system's potential energy by moving out of the zigzag plane and wind around the trap axis.
The local order parameter $\phi_k$ then remains complex,
\begin{equation}
\phi_k = (-1)^k (x_k+i y_k) = |\phi_k| e^{i\theta_k},
\end{equation}
with the angle $\theta_k$ being measured relative to the $x$ axis.  

Before planarization, the helix is rotated about the $z$ axis by an angle $\beta$, which we define such that the local order parameter transforms as
\begin{equation}
    \phi_k(\beta)=e^{-i\beta}\phi_k .
\end{equation}

The initial domain pattern encoded by the rotated helix is obtained from
its projection onto the fixed $x$ direction,
\begin{equation}
    \sigma_k^{i}(\beta) \equiv
    \text{sgn}\left[\operatorname{Re}\phi_k(\beta)\right]=
    \text{sgn}\left[|\phi_k|\cos(\theta_k-\beta)\right].
    \label{eq:rotated_projected_order_parameter}
\end{equation}

Sign changes of $\sigma_k^{i}(\beta)$ identify the projected helical nodes and hence the positions at which planar domain walls are imprinted. Their number and approximate axial positions are therefore encoded in the initial helix and can be tuned through the helix rotation angle $\beta$.

The planarization quench suppresses the radial component $y$ and maps the complex $S^1$ order parameter onto the real $Z_2$ order parameter in the $x$-$z$ plane.
After planarization, the corresponding domain is evaluated
directly in the fixed $x$-$z$ plane,
\begin{equation}
    \sigma_k^{f}
    =
    \operatorname{sgn}\!\left[(-1)^k x_k^{f}\right],
\end{equation}
with $x_k^f$ denoting the $x$ position of ion $k$ once the crystal reaches the planar state. 

To successfully imprint the defects into the planar phase, the projected domain pattern needs to be preserved during planarization. We quantify this by defining a domain fidelity
\begin{equation}
F_\sigma =
\frac{1}{N_{\rm zz}}
\sum_{k\in{\rm zz}}
\delta_{\sigma_k^{\rm i},\sigma_k^{\rm f}}\, .
\end{equation}
The sum is restricted to the number of ions in the zigzag region $N_{\rm zz}$, excluding ions with negligible radial displacement in the nearly linear outer regions. For $F_\sigma=1$, the planarization preserves the full projected domain pattern and all projected helical nodes are converted into planar kinks. For $F_\sigma<1$, parts of the initially encoded pattern reorganize during planarization, so that projected kinks can be lost.

The planarization quench can be separated into two limiting regimes:
\\
\paragraph*{Adiabatic planarization.}
In the adiabatic regime, the crystal is given sufficient time to follow a locally stable configuration during planarization. Kinks are generated if the helical state remains on a finite-winding branch and the corresponding domain fidelity $F_\sigma$ is close to $1$. In that case, the projected helical nodes are converted into planar domain walls.

In contrast to the reversible linear-to-zigzag, the helix-to-zigzag planarization can involve metastable finite-winding branches so that the system does not retrace the defect-free zigzag preparation path but rather keeps an imprinted domain pattern which is then converted into planar defects. 

The specific outcome of an adiabatic planarization is determined by the
geometry of the initial helical state. In general, a finite ion helix does
not have a radially homogeneous charge distribution, but
exhibits a principal axis associated with the lowest projected ion density in the radial plane. During a sufficiently slow planarization ramp, the helix can rotate within the trap potential such that this principal axis aligns with the quench direction. In this case, for a given helix, the projected kink pattern is independent of the chosen planarization basis.

In our simulations, we approach planarization in the adiabatic regime by changing the transverse confinement in small increments. After each increment, damped zero-temperature dynamics is continued until the kinetic energy and the winding pattern remain stationary for at least $\SI{100}{\micro\second}$. This allows the crystal to follow a locally stable or metastable configuration branch during the planarization simulations. We find, that selected precursor helices can be converted into finite-kink states by adiabatic planarization. However, the set of initial helical configurations that lead to stable single- or multi-kink states is substantially enlarged when the quench rate is used as an additional control parameter.
\\
\paragraph*{Finite-rate planarization.}
In the finite-rate regime, the planarization is performed faster than the relaxation processes that would otherwise modify the projected sign pattern. This regime is useful when an adiabatic planarization does not produce the desired kink configuration. If the currently encoded sign pattern is preserved during the finite-rate projection, $F=1$, the helical nodes are again deterministically converted into planar kinks.

The finite-rate quench therefore does not play the same role as in Kibble--Zurek defect formation, where it determines the size of the stochastically formed domains. Instead, it freezes in a pre-existing projected sign pattern before the helix can relax to a different planar outcome. 

In this regime, the planarization direction becomes a direct control parameter. Changing the radial direction of the quench changes the projection pattern of the order parameter onto the final planar $Z_2$ manifold. Together with the trap aspect ratio $\alpha$, this provides control over the number and positions of the projected nodes and therefore over the resulting kink configuration.

The minimal quench rate required to obtain $F=1$ is not universal. It depends on the geometry of the precursor helix, in particular on the axial separation of neighboring projected nodes and on their distance from the crystal boundaries. The quench must be fast enough to suppress three-dimensional reordering before the planar domain pattern is established.

While an in-depth analysis of the minimum quench-rate dependence on $\alpha$ and $N$ is beyond the scope of this work, we find that for all configurations studied in Sec.~\ref{sec:results_defect_formation}, quench times on the order of
\begin{equation}
t_Q \sim \SI{10}{\micro\second}
\end{equation}
are sufficient to obtain $F=1$ when ramping the radial frequency from $\omega^\mathrm{i}_{y'}=\omega_{x'}$ to $\omega^\mathrm{f}_{y'}=1.5\,\omega_{x'}$, corresponding to a quench rate of approximately
\begin{equation}
\gamma_Q \sim 2.3\,\omega_z^2 .
\end{equation}
This ensures that the helical nodes are converted into planar domain walls for the configurations considered here.

Once the crystal has reached the planar phase, the defects then evolve under the usual planar kink dynamics: kink--antikink pairs can attract and annihilate, and defects can be lost at the boundaries of the zigzag region. The stability of the resulting multi-kink states is therefore governed by the kink--kink interaction and by the Peierls--Nabarro landscape, which are analyzed in Sec.~\ref{sec:kink_interaction} and Sec.~\ref{sec:post_selection}.

In our simulations, we generally observe an increased survivability of the defects in the final planar crystal compared to conventional finite-rate linear-to-zigzag quenches, which we attribute to the direction in which kinetic energy is injected during the quench. In conventional linear-to-zigzag quenches, the quench injects kinetic energy in the same plane in which the subsequent kink dynamics evolve. In the deterministic protocol, by contrast, the finite-rate quench suppresses the transverse direction orthogonal to the final crystal plane. The injected energy therefore couples only indirectly to the planar kink degrees of freedom, which can increase the survivability of the formed defects.

\section{Methods}
\label{sec:methods}
\subsection{Molecular dynamics simulations}
We model the ion motion with a Langevin equation that includes both damping and stochastic forces:
\begin{equation}
m_i\frac{d^2\vec r_i}{dt^2}
=
-\frac{d}{d\vec r_i}\mathcal{V}
- m_i\eta\frac{d\vec r_i}{dt}
+ \vec \epsilon_i(t).
\end{equation}
Here, the damping term with coefficient~$\eta$ and the noise term~$\vec{\epsilon}_i(t)$ describe the Brownian dynamics of the ions. 
Doppler-cooling damping and stochastic forces $\varepsilon(t)$, are modeled by the fluctuation-dissipation relation
\begin{equation}
\langle \varepsilon_{\alpha j}(t)\, \varepsilon_{\beta i}(t')\rangle = 2\eta k_B T \delta_{\alpha \beta} \delta_{i j}(t-t')\, ,
\end{equation}
with $\alpha, \beta = x,y,z$~\cite{PykaDefectFormation2013, kiethe_Aubrytype_2017a}. In the experiment, these fluctuations arise from photon absorption and spontaneous emission during laser cooling. For Doppler cooling, the maximum friction coefficient is approximately ${m_i \eta \approx \hbar \pi^2 / \lambda^2}$ and is typically in the range of \SIrange[print-unity-mantissa=false]{e-21}{e-20}{\kilogram\per\second} depending on the ion species~\cite{PykaDefectFormation2013}.
Although the results presented in this paper were obtained at zero temperature, their robustness was verified using finite-temperature simulations at temperatures of order $\SI{1}{\milli\kelvin}$ and with a friction coefficient of $\eta = \qty{1e-20}{\kilogram\per\second}$. Before analysis, the system was evolved for several damping timescales, $\eta^{-1}$, and proper thermalization was verified by monitoring the kinetic energy. 
The integration time step is chosen such that the shortest relevant secular oscillation period is resolved by at least 100 integration steps, which is sufficient to avoid significant numerical inaccuracies. The reliability of our simulations was previously verified by comparisons with experimental data~\cite{ruffert_DomainFormationStructural_2024c}. All simulation results presented here use singly charged ytterbium ions Yb$^+$ with mass $172\,\mathrm{amu}$ and a fixed axial secular frequency of $\omega_z/2\pi = \SI{25}{\kilo\hertz}$.

\subsection{Electrode geometry and control of rotation and quenches}
\label{subsec:electrode_control}
\begin{figure}
  \includegraphics[width=\columnwidth]{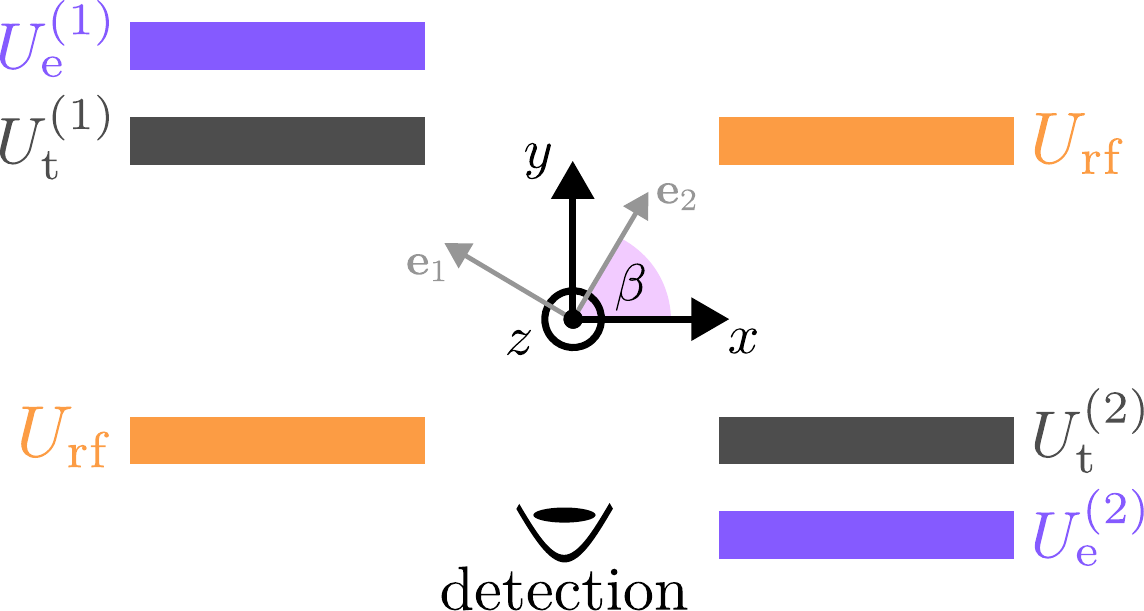}
  \caption{\justifying Schematic radial cross section of the wafer-stack trap~\cite{Herschbach_LinearPaulTrap_2012}. Two rf blades ($U_{\text{rf}}$) provide the radial pseudopotential confinement. Axial confinement along $z$ is achieved by applying voltages to electrodes of neighboring segments. Each rf blade is accompanied by two additional dc electrodes $U_{\text{t}}^{(1,2)}$ and $U_{\text{e}}^{(1,2)}$ that provide two independent degrees of freedom in the radial $(x,y)$ plane. The direction of detection is along $y$. The gray vectors $\mathbf e_1$ and $\mathbf e_2$ indicate the principal axes of the radial curvature matrix. Their controlled rotation is used to rotate the precursor helix by an angle $\beta$ before the crystal is planarized along $y$.}
  \label{fig:trap_schematic}
\end{figure}
Although all results presented in this work are obtained from molecular-dynamics simulations, the simulated rotation and planarization operations can in principle be implemented through experimentally accessible trap controls. We therefore briefly outline this mapping here, while a more detailed relation between electrode voltages and radial curvatures is given in App.~\ref{app:electrode_control}.

In the simulations, we use the fixed laboratory coordinate system introduced in Sec.~\ref{subsec:ion_coulomb_crystals}, where $z$ is the axial trap direction, $y$ is aligned with the detection line of sight, and the observable crystal plane is the $x$-$z$ plane. Planarization is implemented by increasing the secular frequency $\omega_y$, while the desired orientation of the precursor helix is prepared by numerically rotating the ion coordinates about the $z$ axis by an angle $\beta$.

To outline how the same operations may be realized experimentally, we consider the segmented linear Paul trap proposed by Herschbach et al.~\cite{Herschbach_LinearPaulTrap_2012}. Figure~\ref{fig:trap_schematic} shows a schematic radial cross section of the electrode geometry. Near the trap center, the radial potential energy in the fixed laboratory basis $(x,y)$ can be written as
\begin{equation}
    V_r(x,y)=\frac{m}{2}
    \begin{pmatrix}x&y\end{pmatrix}
    \mathbf{K}
    \begin{pmatrix}x\\y\end{pmatrix},
\end{equation}
where $\mathbf{K}$ is the mass-normalized radial curvature matrix
\begin{equation}
    \mathbf{K}
    \equiv
    \begin{pmatrix}
        K_{xx}&K_{xy}\\
        K_{xy}&K_{yy}
    \end{pmatrix}.
\end{equation}
For the experimental implementation, we decompose this matrix as
\begin{equation}
    \mathbf{K}
    =
    \mathbf{K}_0+\frac{Q}{m}\mathbf{H}_{\mathrm{dc}}.
\end{equation}
Here, $\mathbf{K}_0$ denotes the baseline radial curvature in the absence of the additional dc control voltages and includes the rf pseudopotential and all static contributions that remain fixed during the protocol. The matrix $\mathbf{H}_{\mathrm{dc}}$ is the Hessian of the additional electrostatic control potential $\Phi_{\mathrm{dc}}$ at the trap center,
\begin{equation}
    \left(\mathbf{H}_{\mathrm{dc}}\right)_{ij}
    =
    \left.\partial_i\partial_j\Phi_{\mathrm{dc}}\right|_{(0,0)},
    \qquad i,j\in\{x,y\}.
\end{equation}
The additional dc control voltages modify the anisotropy and off-diagonal components of $\mathbf{K}$ and thereby control the orientation of its radial principal axes. The eigenvalues of the complete matrix $\mathbf{K}$ correspond to the squared radial secular frequencies represented directly in the simulations.

The two segmented dc electrode pairs $U_{\mathrm{t}}^{(1,2)}$ and $U_{\mathrm{e}}^{(1,2)}$ provide two independent quadrupolar controls in the radial plane. Diagonally opposite stacks are driven identically,
\begin{equation}
    U_{\mathrm{t}}^{(1)}=U_{\mathrm{t}}^{(2)}\equiv V_t,
    \qquad
    U_{\mathrm{e}}^{(1)}=U_{\mathrm{e}}^{(2)}\equiv V_e,
\end{equation}
thereby suppressing dipole fields and preventing a static displacement of the crystal. Linear combinations of $V_t$ and $V_e$ modify the diagonal anisotropy $K_{xx}-K_{yy}$ and the shear component $K_{xy}$ of the complete radial curvature matrix $\mathbf{K}$, and hence control the orientation of the radial principal axes $\mathbf{e}_1$ and $\mathbf{e}_2$ shown in Fig.~\ref{fig:trap_schematic}.

After preparing a Coulomb helix from a well-ordered zigzag structure as described in Sec.~\ref{subsec:phase_transitions_order_param}, a weak radial anisotropy pins the rotational degree of freedom of the helix. The anisotropic contribution may be rotated according to
\begin{equation}
    \mathbf{K}_{\mathrm{rot}}(\beta)
    =
    \frac{\delta K_{\mathrm{rot}}}{2}
    \begin{pmatrix}
        \cos(2\beta)&\sin(2\beta)\\
        \sin(2\beta)&-\cos(2\beta)
    \end{pmatrix},
    \label{eq:main_rotation_curvature}
\end{equation}
with $\delta K_{\mathrm{rot}}$ characterizing the strength of the anisotropy. The angle $\beta$ specifies the orientation of the corresponding principal axes relative to the fixed laboratory axes. The helix follows the principal-axis frame due to its generally inhomogeneous radial charge distribution. The chosen rotation angle determines the projection of the helical nodes onto the $x$-$z$ detection plane and therefore selects the domain pattern from which the planar kinks are formed.

Once the desired angle $\beta$ has been reached, the rotated pinning anisotropy is reduced while the confinement along the fixed $y$ direction is increased. Conceptually, the complete radial curvature matrix during this quench can be written as
\begin{equation}
    \mathbf{K}(t)=K_0(t)\mathbf{I}+a(t)\mathbf{K}_{\mathrm{rot}}(\beta)+\frac{q(t)}{2}
    \begin{pmatrix}
        -1&0\\
        0&1
    \end{pmatrix},
    \label{eq:main_rotation_quench_handover}
\end{equation}
where $K_0(t)$ denotes the isotropic common confinement of the radial curvature, $a(t)$ gives the amplitude of the radial pinning anisotropy and decreases from unity to zero, and $q(t)$ denotes the additional time-dependent curvature applied along $y$ and increases from zero to its final value during the planarization quench. The final potential therefore has its strong principal axis along the fixed detection direction $y$, confining the crystal to the observable $x$-$z$ plane. Further details on the voltage control and the combined rotation--quench protocol are given in App.~\ref{app:electrode_control}.

\section{Results}
\subsection{Deterministic defect formation}
\label{sec:results_defect_formation}
\begin{figure*}
    \includegraphics[width=\textwidth]{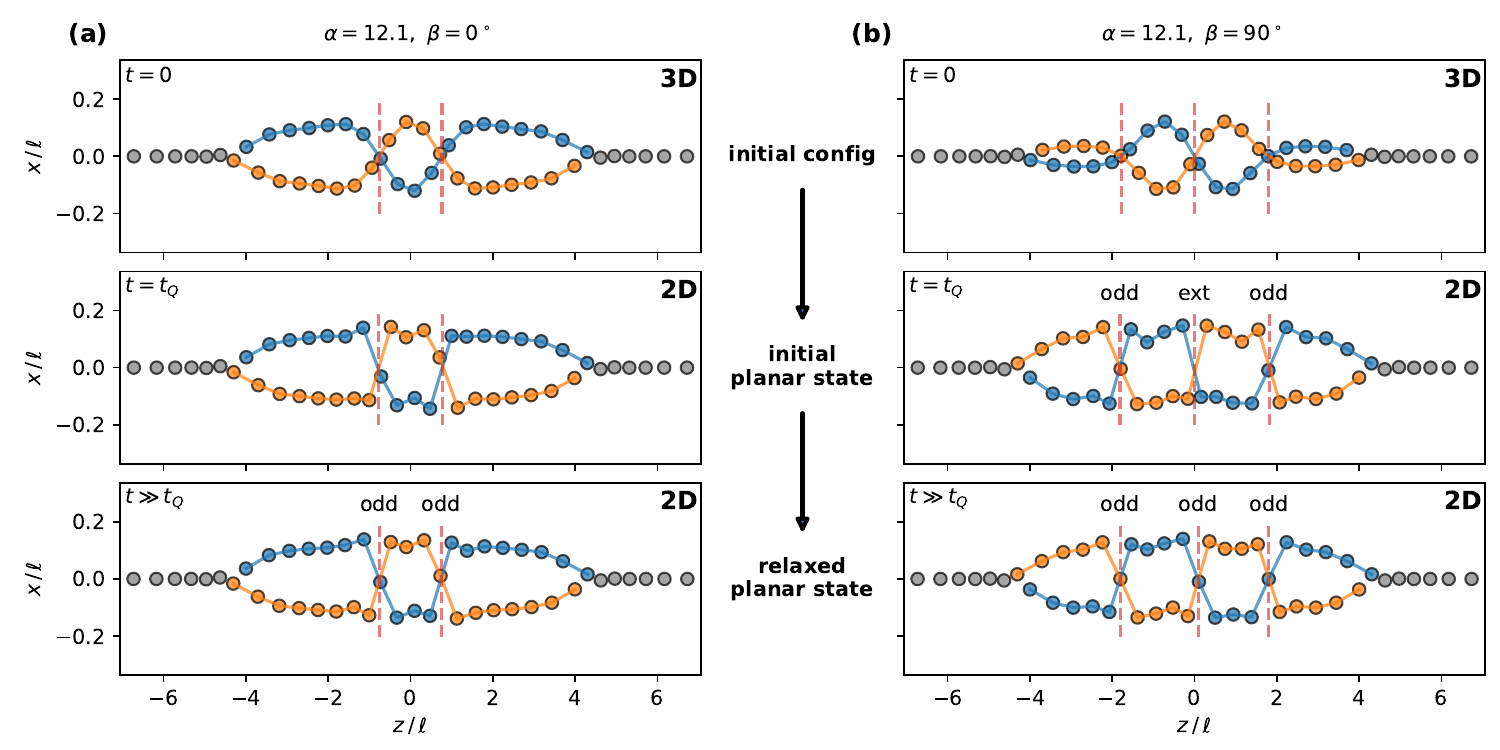}
    \caption{\justifying Deterministic kink formation using a 50-ion crystal forming a single-turn double helix, projected onto the $z$–$x$ plane at aspect ratio $\alpha=12.1$ for two different angles of rotation $\beta$. The quench along the $y$-direction starts at $t=0$ and relative times are indicated in the top left of each frame. Ion positions are highlighted in orange and blue to distinguish the two helical arms. At $t=t_Q$, the crystal has first reached the planar state, while $t \gg t_Q$ shows the final relaxed state. \textbf{(a)} At ${\beta=0^\circ}$ two defects with topological charge of $q_i=+1,-1$ form at the projected nodes at $t=t_Q$. Upon relaxation, both defects evolve into odd-type kinks. \textbf{(b)} At ${\beta=90^\circ}$, three projected nodes are planarized into domain walls (with topological charge $q_i=+1,-1,+1$) which relax into three odd-type defects.}
    \label{fig:N_50_quench_1_winding}
\end{figure*}
\begin{figure*}
    \includegraphics[width=\textwidth]{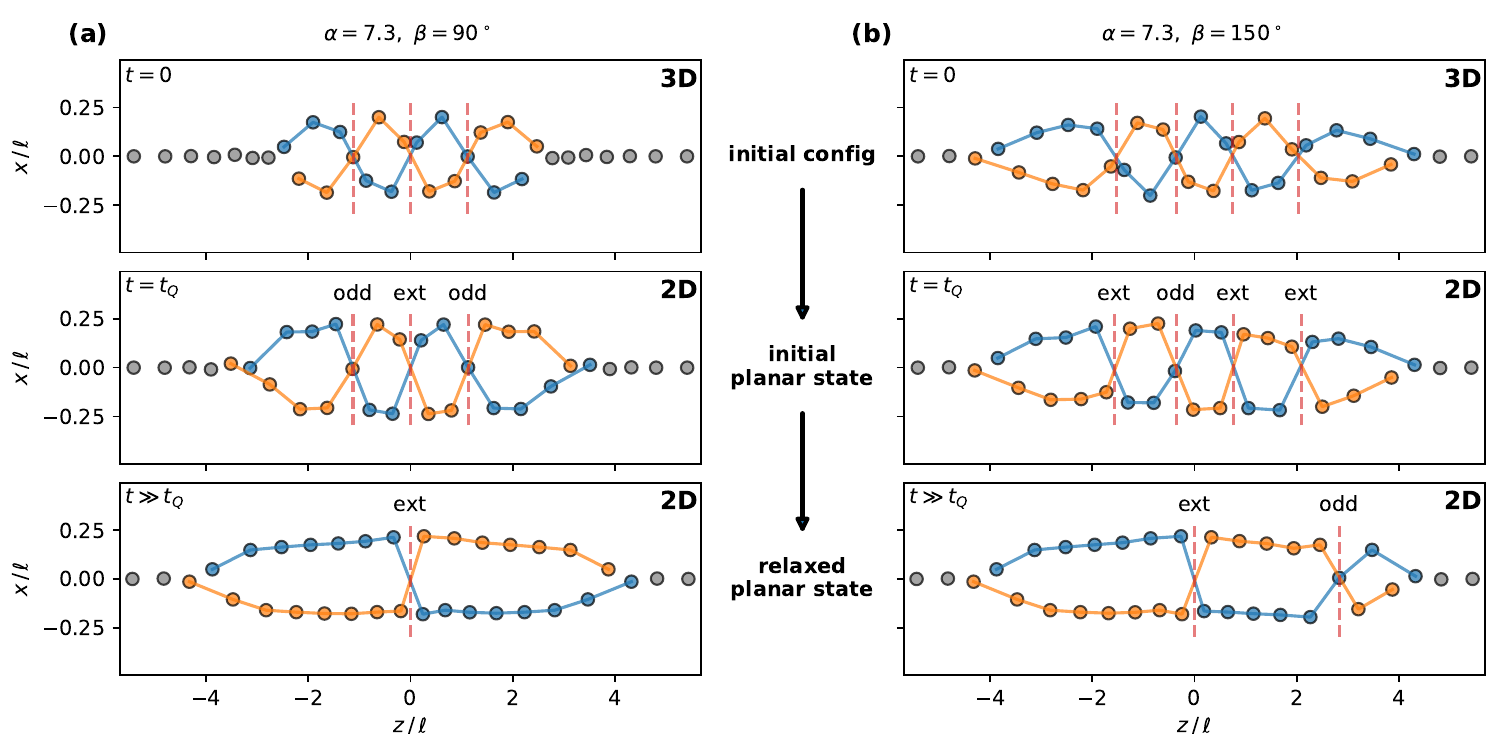}
    \caption{\justifying Deterministic kink formation using a 34-ion crystal forming a two-turn double helix, projected onto the $z$–$x$ plane at aspect ratio $\alpha=7.3$ for two different angles of rotation $\beta$. The quench along the $y$-direction starts at $t=0$ and relative times are indicated in the top left of each frame. Ion positions are highlighted in orange and blue two distinguish the two helical arms. At $t=t_Q$, the crystal has first reached the planar state, while $t \gg t_Q$ shows the final relaxed state. \textbf{(a)} At ${\beta=90^\circ}$ three nodes are projected into three domain walls with topological charge $q_i=+1,-1,+1$. Subsequent annihilation of a kink–antikink pair leaves one extended kink in the lower chain. \textbf{(b)} For ${\beta=150^\circ}$ four nodes are projected into four domain walls at the nodes $z$ position. During relaxation, the inner defect pair annihilates, while the remaining two slightly shift their position in the effective lattice potential. The right defect converts into an odd kink near the edge, while the left kink remains extended. The final state contains two stable defects with charges $q_i=+1,-1$. Running this quench in the adiabatic limit results in the annihilation of all 4 defects.}
    \label{fig:N_34_quench_2_windings}
\end{figure*}

We demonstrate the preparation of single- and multi-kink states by planarization of a given 3D precursor Coulomb helix. We show that the number and approximate axial positions of defects in the planar phase are predictable and reproducible.

We begin with the preparation step of our protocol outlined in Sec.~\ref{subsec:defect_engineering}, by initializing the ions in a defect-free planar zigzag in the $x$-$z$ plane. The system is then brought into radial symmetry by reducing $\omega_y$ until ${\omega_x=\omega_y}$, and the radial confinement is further reduced until a stable helix with the desired winding number is formed. 
Next, the helix is rotated about the $z$-axes by an angle $\beta$, changing the projected domain pattern in the $x$-$z$ plane before planarizing the crystal by increasing $\omega_y$. Specifically, our planarization quench increases the initial $\omega^i_{y}$ to the final ${\omega^f_{y}=1.5\;\omega_{x}}$ over a quench time of $t_Q=\SI{10}{\micro\second}$. This sufficiently planarizes the crystal in the $x$--$z$ plane while preserving the domain fidelity $F_\sigma$ as described in Sec.~\ref{subsec:defect_engineering}. 

In the following, we consider two configurations: ${\alpha=12.1}$, corresponding to a single-turn helix with ${N=50}$, and ${\alpha=7.3}$, corresponding to a two-turn helix with ${N=34}$. Both systems were simulated for temperatures of up to $T=\SI{1}{\milli\kelvin}$, without changing the outcome of the quench, however the configurations shown in Fig.~\ref{fig:N_50_quench_1_winding} and Fig.~\ref{fig:N_50_quench_1_winding} were obtained at $T=0$.


To show the effect of the chosen rotation angle $\beta$ on the produced defect number and positions, the planarization quench is carried out using two different values of $\beta$.

Figure~\ref{fig:N_50_quench_1_winding} shows the planarization quench of the $N=50$ helix with a winding number of $\sim1$ projected onto the $x$-$z$ plane. The two columns (a) and (b) correspond to $\beta=0^\circ$ and $\beta=90^\circ$, respectively and the quench progression is read from top to bottom with the relative time being indicated in the top left corner of each frame. The top row shows the initial 3D configuration in the projected frame, the middle row depicts the moment at which the crystal first reaches the planar phase but has not been equilibrated and the bottom row ($t\gg t_Q$) shows the final relaxed configurations. Ions in the linear phase are colored gray, while the remaining ions are colored blue and orange, alternating along $z$ to visualize the two helical arms. Red dashed lines mark the node positions along $z$.

For $\beta=0^\circ$ (see Fig.~\ref{fig:N_50_quench_1_winding}(a)) the projection exhibits two nodes. At ${t = t_Q}$ two defects have formed near these nodes and relax into a pinned kink--antikink pair at equidistant locations about the trap center. Although opposite topological charges favor annihilation, both defects remain stable due to local trapping minima in the lattice potential. For $\beta=90^\circ$ (Fig.~\ref{fig:N_50_quench_1_winding}(b)) three nodes occur, and three defects with alternating charge form at $t\simeq t_Q$. After equilibration all three remain stable pinned at their initial positions. All kinks are of odd-type as a consequence of the chosen $\alpha$ (see Fig.~\ref{fig:alpha_vs_Nions}). For this system in particular where the nodes are separated by multiple lattice constants, the identical final state is also realized in the adiabatic limit.

We apply the same procedure to an $N=34$ crystal forming a two-turn helix at $\alpha=7.3$ (see Fig.~\ref{fig:N_34_quench_2_windings}). Panels (a) and (b) correspond to $\beta=90^\circ$ and $\beta=150^\circ$, respectively. In Fig.~\ref{fig:N_34_quench_2_windings}(a), three defects form near $t\simeq t_Q$ with alternating topological charge. The central kink--antikink pair annihilates during relaxation, leaving a single extended kink near the center. The quench can be performed in the adiabatic limit without changing the final configuration, as the odd number of nodes in the extended-regime will reliably result in a single centralized when thermal energies are sufficiently low to prevent losses during planarization. 
In Fig.~\ref{fig:N_34_quench_2_windings}(b), four defects form at $t\simeq t_Q$ with alternating charges. The two central defects annihilate, while the outer pair separates and converts into different defect types: the left defect relaxes into an extended kink near the center, whereas the right one converts into an odd kink closer to the crystal edge. 
In this case, a quench in the adiabatic limit results in a defect-free state. The finite-quench rate is therefore required to keep the domain fidelity $F_\sigma\sim1$ and result in the multi-defect configuration shown. 

The examples given illustrate how the number of helix nodes in the projected plane determines the number of defect nucleation sites, while their axial positions are inherited from the node positions. Choosing the rotation angle $\beta$ of the precursor helix provides a direct way to tune the defect number and placement, while the ratio of the trapping frequencies $\alpha$ enables additional control over the helix winding number and the defect-regime. Once the planar phase is reached, subsequent relaxation can cause closely spaced kink--antikink pairs to annihilate due to their alternating topological charge. In the examples considered here, quenches in the adiabatic limit are possible, provided the kink--antikink interaction is suppressed by sufficiently deep local minima of the effective lattice potential, or an odd number of nodes which enforces a nonzero net topological charge after the quench.
While the planarization quench determines the approximate positions and number of defects, the annihilation dynamics in the planar phase are governed by the 2D kink--kink interactions which we will discuss next.

\subsection{Kink--kink interaction and PN-mediated annihilation thresholds}
\label{sec:kink_interaction}
\begin{figure*}[t] 
  \begin{subfigure}{0.5\textwidth}
    \includegraphics[width=\linewidth]{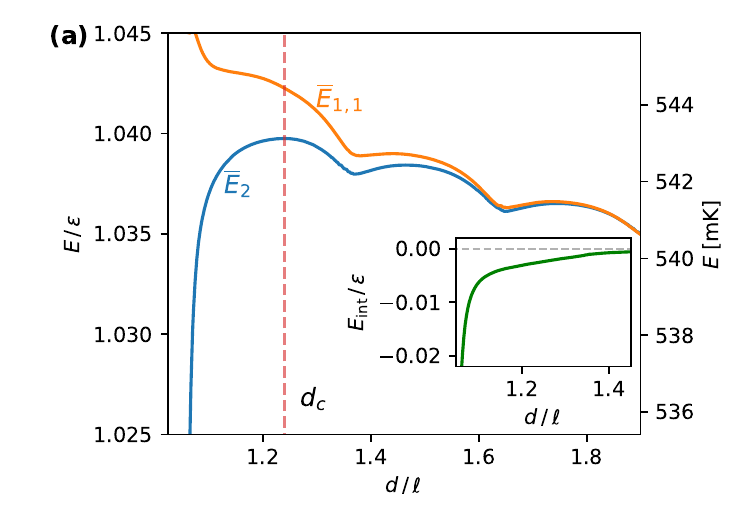}
  \end{subfigure}%
  \begin{subfigure}{0.5\textwidth}
    \includegraphics[width=\linewidth]{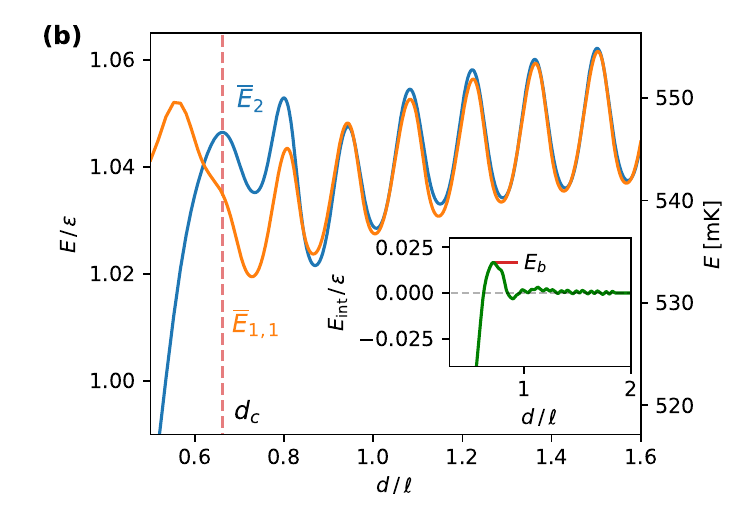}
  \end{subfigure}
  \caption{\justifying Two-kink PN energy $\overline{E}_2$ compared to the energy $\overline{E}_{1,1}$ constructed from single-kink excitations for extended kinks \textbf{(a)} and odd kinks \textbf{(b)}, shown as a function of the kink separation $d$. The data were evaluated for $N=100$ ($101$)  ions at $\alpha=18.0$ ($24.6$) for extended (odd) kinks. The energy of the kink-free configuration was subtracted from both curves. Insets show the corresponding interaction energy. The secondary y-axis gives the equivalent energy scale in $\si{\milli\kelvin}$ for the chosen ion species $^{172}$Yb$^+$ and reference frequency $\omega_z / 2 \pi=\SI{25}{\kilo\hertz}$. In panel \textbf{(a)}, the red dashed line indicates the critical distance $d_c$, below which the two kinks annihilate. In panel \textbf{(b)}, the red, solid line highlights the odd-kink interaction barrier $E_b$, defined as the positive local maximum of $E_\mathrm{int}(d)$ prior to annihilation.}
  \label{fig:kink_kink_interaction_main}
\end{figure*}

\begin{figure*}[t] 
\includegraphics[width=\linewidth]{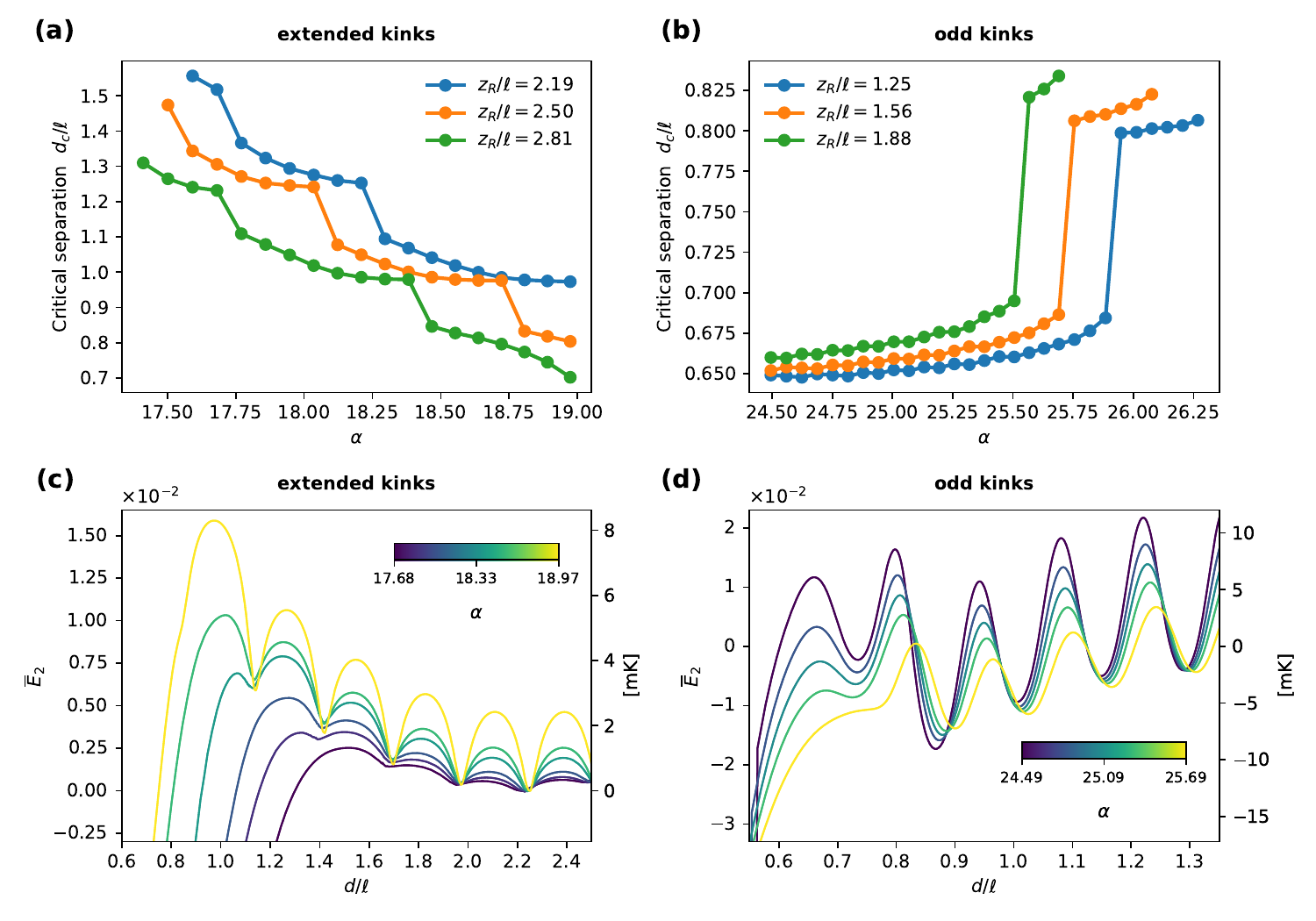}
\caption{\justifying Critical separation $d_c$ below which the two kinks annihilate as a function of $\alpha$ for three initial positions of the right kink, shown for extended kinks \textbf{(a)} in an $N=100$ crystal and odd kinks \textbf{(b)} in an $N=101$ crystal. For extended kinks, $d_c(\alpha)$ decreases with increasing $\alpha$, consistent with stronger pinning at larger radial confinement. For odd kinks, $d_c(\alpha)$ increases with $\alpha$. The step-like structure reflects the appearance (extended) or disappearance (odd) of local pinning maxima as $\alpha$ is increased, with the step size being set by the PN modulation period $\lambda_\text{PN}$. Panel \textbf{(c)} and \textbf{(d)} show the Two-kink PN energy $\overline{E}_{2}$ with respect to the position of the kink-kink distance $d$ for different trapping anisotropies $\alpha$. \textbf{(c)} Extended-kink regime with the right kink at $z_R/\ell = 2.19$. The two-kink potential develops a new local maximum at smaller distances with increasing $\alpha$. \textbf{(d)} Odd kink regime with a right kink at $z_R/\ell = 1.25$. Increasing $\alpha$ shows the gradual disappearance of the first local maximum. The sharp drop of $E_2$ marks the annihilation of both defects.}

\label{fig:critical_separation_vs_alpha}
\end{figure*}
\begin{figure}[t]
  \centering
  \includegraphics[width=\linewidth]{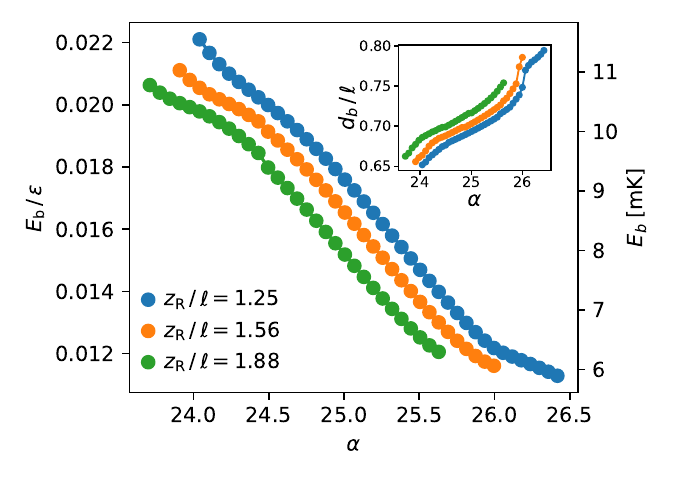}
  \caption{\justifying Interaction barrier $E_b$ of odd-kink pairs as a function of the trap anisotropy $\alpha$ for different initial positions of the right kink $z_R$. The secondary y-axis gives the equivalent energy scale in $\si{\milli\kelvin}$ for the chosen ion species $^{172}$Yb$^+$ and reference frequency $\omega_z / 2 \pi=\SI{25}{\kilo\hertz}$. The barrier decreases monotonically with increasing $\alpha$, demonstrating tunability of the effective kink--kink interaction. Inset: Corresponding barrier position $d_b$ (in units of the characteristic length $\ell$) as a function of $\alpha$. The slight waviness visible in $E_b(\alpha)$ correlates with a non-uniform shift of $d_b(\alpha)$, reflecting the discrete Peierls--Nabarro landscape and lattice-registry effects in the finite crystal.}
  \label{fig:odd_kink_energy_barrier_vs_alpha}
\end{figure}
The interaction of two kinks in trapped-ion crystals has previously been investigated experimentally and theoretically by Landa et al.~\cite{landa_StructureDynamicsBifurcations_2013}. In particular, they demonstrated interaction-stabilized two-kink configurations and calculated a finite barrier against annihilation for a specific trapping configuration. Here, we extend this picture by extracting the effective mutual interaction energy $\delta E_\text{int}(d)$ between the defects from the full two-kink Peierls--Nabarro (PN) potential for both odd and extended defects. We study its dependence on trap anisotropy and absolute defect position and analyze the smallest separation at which a metastable two-kink configuration can still exist before neighboring defects annihilate.

The total energy of a crystal containing a single kink at position $X$ can be written as
\begin{equation}
    E_1(X) = E_0 + \delta E_\text{kink}(X),
    \label{eq:single_kink_energy}
\end{equation}
where $E_0$ is the potential energy of the defect-free crystal and $\delta E_\text{kink}(X)$ is the excess energy associated with the kink-induced lattice distortion.

From the two single-kink PN potentials, we construct the energy
\begin{equation}
\begin{split}
    E_{1,1}(X_1,X_2) &\equiv E_1(X_1)+E_1(X_2)-E_0 \\
    &= E_0+\delta E_\text{kink}(X_1)+\delta E_\text{kink}(X_2),
\end{split}
\label{eq:single_kink_constructed_energy}
\end{equation}
which is the energy of the superposition of the two independently calculated single-kink configurations without the kink--kink interaction or additional energy contribution arising from lattice distortions.

The two-kink PN potential obtained from a direct minimization of a crystal containing both kinks is decomposed as
\begin{equation}
\begin{split}
    E_2(X_1,X_2)&\approx E_{1,1}(X_1,X_2)+\delta E_\text{int}(d)\\&\quad+\delta E_\text{shift}(X_1,X_2),
    \label{eq:two_kink_energy}
\end{split}
\end{equation}
where $d=|X_2-X_1|$. Here, $\delta E_\text{int}(d)$ denotes the kink--kink interaction energy, while $\delta E_\text{shift}(X_1,X_2)$ describes a slowly varying background contribution associated with the additional lattice distortion in the two-kink crystal.

To estimate $\delta E_\text{shift}(X_1,X_2)$, the right kink is initialized at $X_2>\lambda_\text{PN}$ and the position $X_1$ of the left kink is varied for $X_1<0$. For distances much larger than the period of the PN energy modulation $d\gg\lambda_\text{PN}$, the direct kink--kink interaction is assumed to be negligible. The difference between the two-kink energy and the constructed two-single-kink energy then provides the energy shift
\begin{equation}
\begin{gathered}
    \delta E_\text{shift}(X_1,X_2)
    \approx E_2(X_1,X_2)-E_{1,1}(X_1,X_2),\\
    X_1<0,\quad d\gg\lambda_\text{PN}.
\end{gathered}
\label{eq:energy_shift}
\end{equation}
Based on the reflection symmetry of the underlying trapping potential, we approximate this background contribution as symmetric about the trap center.
The kink--kink interaction energy is then obtained by subtracting the resulting background curve from the complete energy difference,
\begin{equation}
\begin{split}
    \delta E_\text{int}(d)&=E_2(X_1,X_2)-E_{1,1}(X_1,X_2)\\&\quad-\delta E_\text{shift}(X_1,X_2).
\end{split}
    \label{eq:interaction_energy}
\end{equation}

We consider a crystal with $N=100$ ($N=101$) ions for extended (odd) kinks and subtract the energy of the kink-free configuration of the calculated potentials,
\begin{align}
    \overline{E}_{1,1} &= E_{1,1}-E_0, &
    \overline{E}_{2} &= E_{2}-E_0.
\label{eq:shifted_energies}
\end{align}
Both quantities are plotted as a function of the separation $d$ for extended kinks in Fig.~\ref{fig:kink_kink_interaction_main}(a) and for odd kinks in Fig.~\ref{fig:kink_kink_interaction_main}(b). The right kink is initialized at a fixed position in the right half of the crystal and is unconstrained, while the left kink is moved towards it in small increments using constrained energy minimization until the defects annihilate and the crystal relaxes to the defect-free ground state. For determining the positions of the kinks in a multi-kink configuration, we use the position definitions for single kinks by Partner et al.~\cite{PartnerDefects2013a} with a slight modification for extended kinks which is explained in more detail in Appendix~\ref{app:multi_kink_position}.

For extended kinks (Fig.~\ref{fig:kink_kink_interaction_main}(a)), $\overline{E}_{1,1}$ closely follows $\overline{E}_2$ down to separations of about $d/\ell \approx 1.6$, while over the full measured range we find $\overline{E}_2 \lesssim \overline{E}_{1,1}$. The sharp drop in $\overline{E}_2$ marks the annihilation event. The extracted interaction energy (inset of Fig.~\ref{fig:kink_kink_interaction_main}(a)) decreases monotonically as the kinks approach each other, while converging to zero for large separations. The interaction remains attractive over the investigated parameter range. The barrier that must be overcome before annihilation is thus not set by the isolated interaction energy $\delta E_\text{int}(d)$, but rather by the global modulation of the PN landscape.

Odd kinks show a qualitatively different behavior. The extracted interaction energy exhibits a local maximum at positive values of $E_\text{int}$ (inset of Fig.~\ref{fig:kink_kink_interaction_main}(b)), indicating a repulsive barrier prior to the final attractive approach and annihilation. This repulsive interaction is intrinsic to the kink--kink interaction in the odd regime and is found consistently for different values of $\alpha$ and initial positions of the right kink. 

In short, Fig.~\ref{fig:kink_kink_interaction_main} shows that the interaction between extended kinks is attractive over the investigated parameter range, whereas odd kinks exhibit an additional short-range repulsive barrier of magnitude $E_b(\alpha)$.

\subsubsection*{$\alpha$-dependence of the annihilation-distance threshold}

Since extended kinks do not provide an intrinsic interaction barrier, we next characterize the annihilation process through an operational metastability threshold. 
We define $d_c$ as the smallest kink separation for which a metastable two-kink branch still exists in the full PN landscape. 
This threshold is governed by the last local PN maxima that separate metastable two-kink configurations from the annihilated, defect-free state. 
Because $\alpha$ changes the PN modulation depth, it affects the separation at which this metastable branch terminates. We extract $d_c$ from $E_2(d)$, which is illustrated by the red dashed lines in Fig.~\ref{fig:kink_kink_interaction_main}. 
For both defect types, we evaluate $d_c(\alpha)$ for three initial right-kink positions, since the finite-crystal PN landscape depends not only on $\alpha$ but also on the absolute defect position. The results are shown in Fig.~\ref{fig:critical_separation_vs_alpha}.

For extended kinks (Fig.~\ref{fig:critical_separation_vs_alpha}(a)), $d_c(\alpha)$ decreases monotonically with $\alpha$. Increasing $\alpha$ strengthens the radial confinement, increases PN modulation amplitudes (see Fig.~\ref{fig:PN_potentials}(b)), and therefore enhances pinning, allowing metastable two-kink states down to smaller separations. Since the PN modulation also grows towards the crystal edges, placing the right kink closer to the crystal edges results in enhanced pinning of the defect and shifts $d_c(\alpha)$ to smaller values. Superimposed on this monotonic decrease, we observe discrete steps in $d_c(\alpha)$. These occur when additional local maxima appear in the two-kink PN potential, enabling extra metastable configurations and shifting $d_c$ by approximately $\lambda_\text{PN}$. The continuous change of the two-kink PN landscape with increasing $\alpha$ is exemplified in Fig.~\ref{fig:critical_separation_vs_alpha} (c) and (d). 

For odd kinks (Fig.~\ref{fig:critical_separation_vs_alpha}(b)), the trend is reversed: increasing $\alpha$ reduces the PN modulation amplitude (see Fig.~\ref{fig:PN_potentials}(a)), and metastable two-kink states are only maintained at larger separations, causing $d_c(\alpha)$ to increase with $\alpha$. Moving the right kink towards the crystal edge has a similar effect, as the PN modulation flattens towards the edges. Within the accessible stability region, a single step in $d_c(\alpha)$ is resolved for all three initial configurations. This discontinuity occurs when the barrier at the critical distance disappears, shifting $d_c$ outward by approximately $\lambda_\text{PN}$. The examinable $\alpha$ range is limited by the regime in which stable odd-kink configurations can be realized.

Comparing the critical annihilation distances to the period of the PN modulation, $\lambda_\text{PN}\approx0.28$ for extended and $\lambda_\text{PN}\approx0.15$ for odd kinks, shows that $d_c$ is on the order of a few periods of $\lambda_\text{PN}$. This underpins the observation that closely spaced nodes of the double helix in the multi-kink preparation tend to annihilate during the 3D$\to$2D quench.

Overall, Fig.~\ref{fig:critical_separation_vs_alpha} shows that the annihilation threshold of a kink pair is not fixed, but can be tuned systematically by $\alpha$ and by the position of the defects in the finite crystal. In this sense, $\alpha$ controls the metastability region of two-kink configurations even in the absence of an intrinsic repulsive barrier, as is the case for extended kinks. Thus, for extended kinks, $d_c(\alpha)$ provides a relevant criterion for multi-kink preparation: projected helix nodes separated by less than $d_c$ are expected to annihilate during relaxation, as no intrinsic repulsive interaction barrier exists that could provide additional prevention for annihilation. For odd kinks, the existence of a positive maximum in $E_\text{int}(d)$ allows a more direct energetic characterization of the interaction, which we discuss next.

\subsubsection{Tunable odd-kink interaction barrier}
While $d_c$ characterizes the metastability threshold in the full PN landscape, odd kinks allow for an additional and more direct interaction measure. 
For this defect type, the extracted interaction energy exhibits a positive local maximum, defining a repulsive short-range barrier $E_b$ that is absent for extended kinks. To analyze the relation $E_b(\alpha)$, we define the barrier height $E_b$ as 
\begin{equation}
    E_b (\alpha)\equiv \max \left[ E_\text{int}(d, \alpha) \right],
\end{equation}
and calculate it as a function of $\alpha$ in the stable odd-kink regime for different initial positions of the right kink. The results are shown in Figure~\ref{fig:odd_kink_energy_barrier_vs_alpha}. 

The barrier height decreases monotonically with increasing $\alpha$, showing that the short-range repulsive part of the odd-kink interaction can be tuned systematically by the anisotropy of the trapping potential. The three curves are vertically offset showing that odd kinks placed further away from the crystal center not only experience a flatter local modulation of the PN potential but as a result also a smaller barrier height for kink annihilation.
Beyond the overall monotonic trend, a weak modulation of $E_b(\alpha)$ is present in all three graphs, noticeable by a change in slope towards the ends of the graphs. A closer analysis shows that this modulation stems from the $\alpha$-dependence of the barrier position $d_b(\alpha)$, at which the interaction energy is evaluated. While the contributions of $E_{1,1}$ and $E_2$ to the energy barrier $E_b$ evolve approximately linearly with $\alpha$, the location of $E_b$ shifts non-uniformly. This behavior arises as a change in $\alpha$ moves the barrier position across a spatially modulated Peierls-–Nabarro energy landscape, thereby probing different local lattice registries relative to the discrete ion positions (see inset of Fig.~\ref{fig:odd_kink_energy_barrier_vs_alpha}). We therefore attribute the observed modulation to finite-size and lattice-registry effects, rather than to an intrinsic characteristic of the kink-kink interaction.
Our findings demonstrate that the short-range repulsive part of the odd-kink interaction can be tuned continuously by the anisotropy of the trap potential $\alpha$.


Overall, these results separate two distinct control mechanisms in the two-kink problem. The critical distance $d_c$ characterizes the metastability and annihilation threshold of kink pairs in the full two-kink PN landscape and is therefore relevant for both defect species. The barrier height $E_b$, by contrast, is specific to odd kinks and quantifies the intrinsic repulsive part of the kink--kink interaction. 
Extended kinks do not exhibit an analogous barrier, since their extracted interaction energy remains attractive over the resolved range. Thus, $\alpha$ tunes the survival threshold of multi-kink configurations through the PN landscape, and, in the odd-kink regime, the intrinsic short-range repulsive interaction.
In the next step of our protocol, we exploit defect-specific stability regions to deliberately destabilize unwanted defects and thereby post-select tailored multi-kink configurations.

\subsection{Post-selection via defect-specific stability regions}
\label{sec:post_selection}
\begin{figure*}
  \includegraphics[width=0.9\textwidth]{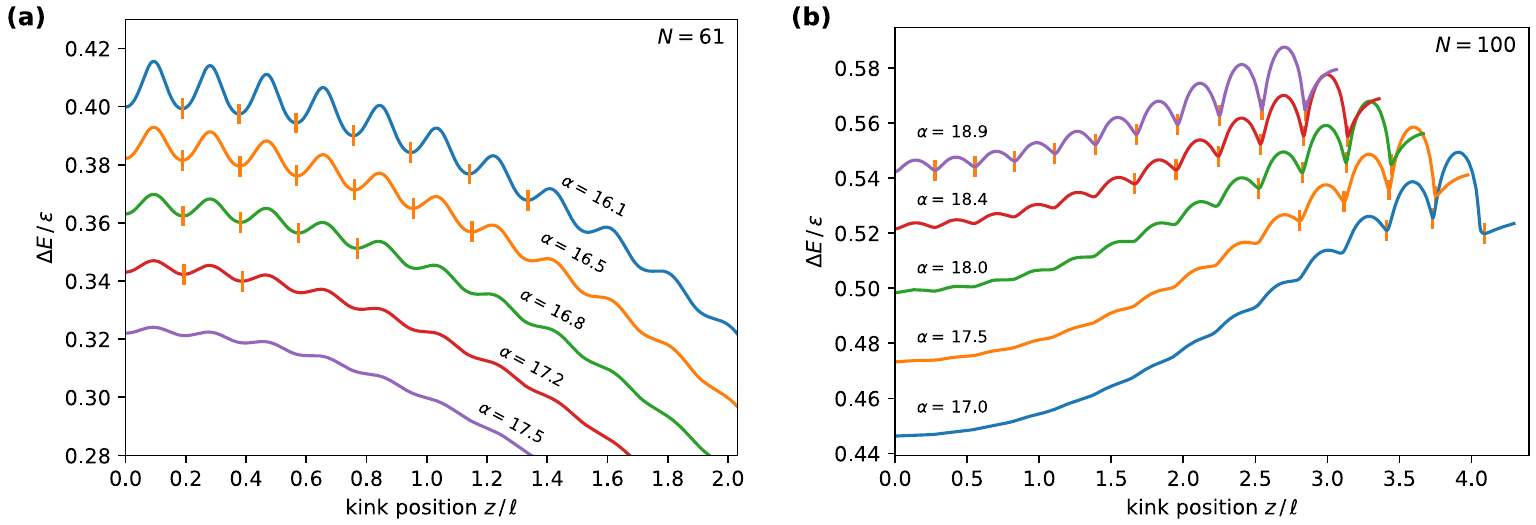}
  \caption{Peierls-Nabarro potential for \textbf{(a)} odd kinks in an $N=61$ and for \textbf{(b)} extended kinks in an $N=100$ 2D ion crystal with varying aspect ratio of the trapping frequencies $\alpha$. Stable kink trapping points based on the threshold given in \eqref{eq:trapping_threshold} are highlighted with orange vertical lines. The PN potential has mirror symmetry with respect to $z=0$, therefore the plotted results are valid in both axial directions.
}
  \label{fig:PN_potentials}
\end{figure*}
After creating a multi-kink state as described in Sec.~\ref{sec:results_defect_formation}, the aspect-ratio of the trapping potential $\alpha$ can be used to destabilize unwanted defects in the crystal to obtain a desired configuration in the experiment. Tuning $\alpha$ drives unwanted defects into regimes where they delocalize and annihilate (either at the boundaries or by kink--antikink interaction), while target defects remain stable. In addition, by sweeping $\alpha$ across kink-regime boundaries, controlled conversions between odd and extended kinks becomes possible (see Appendix~\ref{app:defect_regimes}). To quantify the spatial range and its dependence on $\alpha$ where kinks remain stable on an experimental energy scale, we calculate the PN-potential (see Sec.~\ref{sec:kink_interaction}) for single kink crystals over a broad range of $\alpha$ and $N$. Our findings act as a post-selection filter in experimental scenarios where a specific multi-kink state is desired.

We calculate the energy difference
\begin{equation}
    \Delta E(z) = E\!\left(\mathbf{Q}\,\big|\,g(\mathbf{Q})=z\right) - E_0,
\end{equation}
where $\mathbf{Q}=(z_1,\ldots,z_N,x_1,\ldots,x_N)$ denotes the ionic coordinates in the planar crystal, $g(\mathbf{Q})$ is the defect-center functional, and $E_0$ is the total potential energy of the corresponding defect-free zigzag crystal.
To classify PN minima as practically stable trapping points, we require that they are separated from neighboring maxima by an energy barrier exceeding a threshold $\Delta E_{\mathrm{th}}$. Motivated by residual energies in experiments~\cite{Berkeland_micromotion_1998}, we adopt a reference value of $1\,\mathrm{mK}$, which corresponds to
\begin{equation}
    \Delta E_{\mathrm{th}} / \varepsilon \simeq 1.9\times10^{-3}
    \label{eq:trapping_threshold}
\end{equation}
in the dimensionless energy units defined in Eq.~\eqref{eq:energy_scale}. Minima fulfilling this threshold are counted as stable trapping sites in our analysis. 
Figure~\ref{fig:PN_potentials} illustrates representative PN potentials for (a) odd and (b) extended kinks, with all minima satisfying the threshold being highlighted.
The respective PN minima define an axial stability region for each defect type and each pair $(N,\alpha)$: within this region, defects can remain pinned at sufficiently low temperature, whereas outside it they delocalize and are expelled or annihilate during relaxation.
In the following we summarize the stability regions for odd and extended kinks and highlight their use for post-selection.

\subsubsection{Odd-kink stability region}
\label{sec:odd_post_selection}
For odd kinks, the PN potential is modulated on the lattice scale but decreases towards the crystal edges due to finite-size inhomogeneity. Consequently, odd kinks are only metastable in local PN minima and are expelled once they approach the boundary region where trapping minima disappear \cite{PartnerDefects2013a, landa_StructureDynamicsBifurcations_2013}.
We characterize the stability region by a single boundary $z_{\max}(N,\alpha)$, defined as the largest axial position for which at least one threshold-stable PN minimum exists:
\begin{equation}
    |z| \le z_{\max}(N,\alpha).
\end{equation}
Here, $z_{\max}$ is extracted from the last stable minimum when scanning the kink position from the crystal center towards the edge. Within the odd-kink regime, increasing $\alpha$ reduces the PN corrugation and causes stable minima to disappear from the outside in, thereby reducing $z_{\max}$ (see Fig.~\ref{fig:PN_potentials} (a)).
The resulting dependence $z_{\max}(N,\alpha)$ for different crystal sizes and alpha ranges as well as phenomenological fits to the extracted boundaries are provided in Appendix~\ref{app:stability_fits}.

\subsubsection{Extended-kink stability region}
The PN potential of an extended kink (see Fig.~\ref{fig:PN_potentials} (b)) exhibits a qualitatively different structure compared to the odd-kink case. 
While it is likewise periodically modulated on the lattice scale, the global PN potential of an extended kink increases towards the crystal edges and displays a modulation period approximately twice that of odd kinks. 
The amplitude of this modulation is controlled by the transverse separation of the two ion chains and therefore depends on the trap anisotropy $\alpha$.

For sufficiently large $\alpha$, the PN potential develops local minima along the axial direction in which extended kinks can be trapped. Upon decreasing $\alpha$, the transverse separation between the two chains increases and the modulation of the PN potential flattens, starting from the center of the crystal. As a consequence, stable trapping minima near $z=0$ disappear first, while local minima 
persist at larger axial positions, close to the crossover into the intermediate kink regime.

This behavior differs from the odd-kink case, where local trapping minima 
become unstable from the outside in. For extended kinks, stable trapping points therefore do not collapse to a single boundary, but instead form a parametrically bounded interval away from the crystal center. We define the stability region of an extended kink as the range
\begin{equation}
    z_\text{min}(N,\alpha) \le |z| \le z_\text{max}(N,\alpha),
\end{equation}
within which the PN potential exhibits local minima that satisfy the trapping threshold \eqref{eq:trapping_threshold}.

We note that the global minimum at $z=0$ is always present, independent of $\alpha$, and is excluded from the present analysis. The full dependence $z_{\max}(N,\alpha)$ with phenomenological fits is provided in Appendix~\ref{app:stability_fits}.

\subsubsection{Operational post-selection of multi-kink crystals}
\label{sec:operational_post_selection}
\begin{figure*}
  \includegraphics[width=0.8\textwidth]{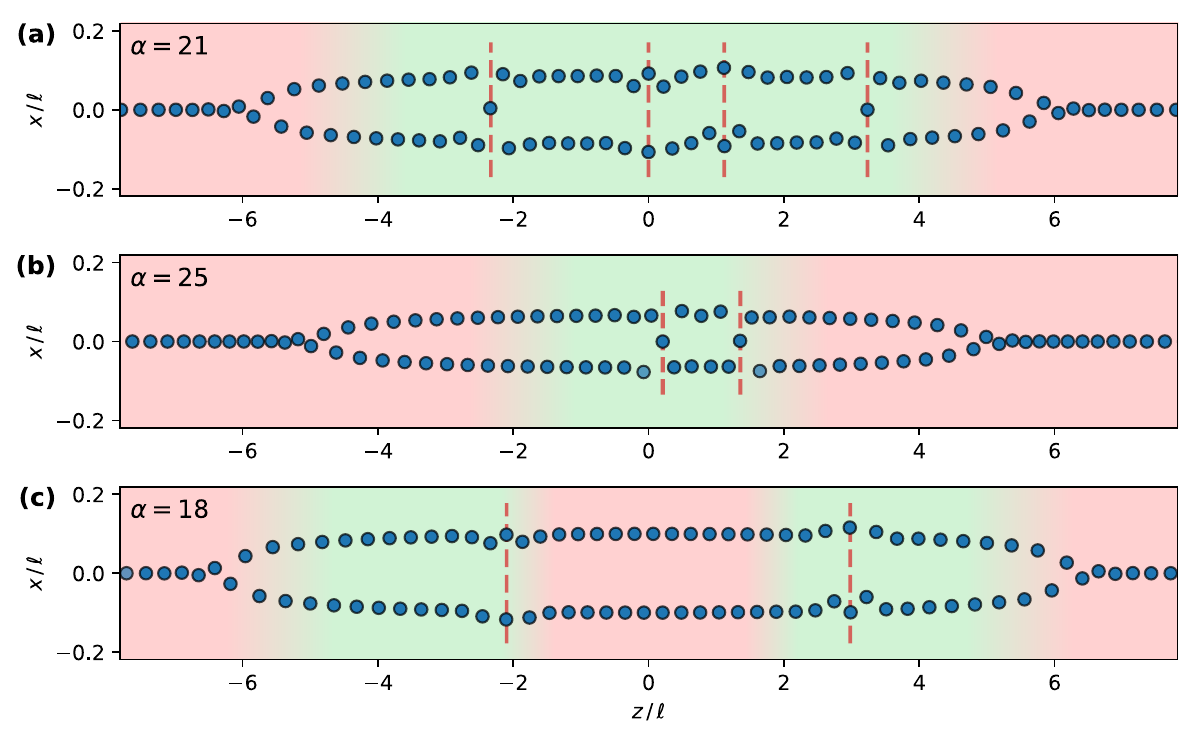}
  \caption{\justifying Example of a post-selection procedure in a multi-kink crystal with $N=100$ ions, created from a {3D$\to$2D} quench (see Sec.~\ref{sec:results_defect_formation}). The initial crystal contains 4 defects (top panel), kink positions are marked with red dashed lines. The color gradient marks stable (green) and unstable (red) trapping regions of the Peierls-–Nabarro potential. \textbf{(b)} An adiabatic quench to $\alpha=25$, destabilizes the outer odd kinks which are lost at the crystal edges, leaving two odd kinks. \textbf{(c)} Alternatively, an adiabatic quench to $\alpha=18$, drives the defects into the extended regime and destabilizes the two inner defects, which annihilate due to attraction between opposite topological charges.}
  \label{fig:multi_kink_selection_scheme}
\end{figure*}
The stability regions provide a direct post-selection tool for the multi-defect states created in Sec.~\ref{sec:results_defect_formation}. Starting from a reproducible quench outcome with defect positions $\{z_i\}$, one can adiabatically tune the trap anisotropy $\alpha$ such that only defects whose positions lie within the corresponding stability region remain pinned, while all others are driven into regimes where the PN trapping minima vanish and the defects delocalize or annihilate. 
Odd kinks are retained by choosing $\alpha$ such that
$|z_i|\le z_{\max}(N,\alpha)$, whereas extended kinks are retained by requiring $z_{\min}(N,\alpha)\le |z_i|\le z_{\max}(N,\alpha)$. By sweeping $\alpha$ across the odd/extended-regime boundary (see Appendix~\ref{app:defect_regimes}), the same procedure can be used to induce controlled defect conversion, enabling the preparation of tailored multi-kink configurations from a single geometric precursor.

Next, we demonstrate the post selection scheme described in Sec.~\ref{sec:post_selection} on an example configuration using $N=100$ ions. After initializing a multi-kink state at an aspect ratio of the trapping frequencies of $\alpha=21$ using the 3D$\to$2D quenching method described in Sec.~\ref{sec:results_defect_formation},  the resulting crystal contains 4 kinks, inhomogeneously distributed along the $z$ axes. Fig.~\ref{fig:multi_kink_selection_scheme} (a) shows the initial crystal configuration. To eliminate the outermost defects, we drive an adiabatic quench to $\alpha=25$, where the outer two kinks are destabilized and are lost at the crystal edges (Fig.~\ref{fig:multi_kink_selection_scheme} (b)). The target $\alpha$ can be estimated by the results presented in Fig.~\ref{fig:odd_kink_range}, which provides the last stable position $z_\text{max} / \ell$ for odd kinks and its dependence on $\alpha$ and particle number $N$. 

Alternatively, we can force the annihilation of the inner two defects by reducing $\alpha$ and transform the odd kinks adiabatically into extended defects. For $\alpha=18$, the axial range of stable pinning sites is reduced, forcing the inner two defects to annihilate, leaving only the outer two defects (Fig.~\ref{fig:multi_kink_selection_scheme} (c)).

Our post-selection scheme demonstrates how the stability regions of odd and extended defects can be exploited to specifically modify the multi-kink crystal using $\alpha$ as an experimentally accessible control knob.

\section{Summary}
In this work, we established a deterministic protocol for preparing and tailoring topological defects in single-species ion Coulomb crystals in the planar zigzag phase. This addresses a central limitation of conventional non-adiabatic linear-to-zigzag quenches, where defect formation is stochastic and subsequent kink motion and annihilation complicate the reproducible preparation of multi-kink states. Our approach instead encodes the desired domain structure in a three-dimensional precursor helix and maps it onto the planar $Z_2$ order-parameter manifold by a controlled planarization quench.

Using numerical simulations, we first demonstrated that the number and approximate axial positions of the resulting defects are determined by the projected nodes of the initial helix. We identified two regimes of deterministic defect preparation. In the adiabatic regime, selected helical states remain on metastable branches during planarization and yield reproducible single- or multi-kink configurations. In the finite-rate regime, sufficiently fast planarization can preserve the initially encoded domain pattern even for precursor states that would reorganize in the adiabatic regime, thereby substantially enlarging the range of helices from which defect states can be prepared. In this regime, the rotation of the helix about the trap axes provides additional control over the number and positions at which the kinks nucleate.

After planarization, the survival and dynamics of the defects are governed by the effective Peierls--Nabarro potential and by the mutual kink--kink interactions. We identified a critical annihilation distance that characterizes the termination of metastable two-kink configurations with respect to the anisotropy of the trapping potential. We further isolate the mutual interaction energy which reveals different interaction dynamics for the two defect species:
While the mutual interaction between extended kinks is attractive over the investigated parameter range, odd kinks exhibit a short-range repulsive energy barrier before the final attractive approach and annihilation. We show that both, the metastability threshold and, for odd kinks, the repulsive interaction barrier can be controlled through the trap anisotropy $\alpha$.

Finally, we introduced a post-selection scheme for multi-kink crystals that exploits the defect-specific stability regions of the Peierls--Nabarro potential. For odd kinks, the stable region is bounded by a single outer edge and widens with decreasing $\alpha$. For extended kinks, stable trapping sites form a spatial interval away from the crystal center, which widens when $\alpha$ is increased.
By adiabatically tuning $\alpha$, kinks in selected regions can be destabilized and removed while others remain pinned. Sweeps across the odd-extended regime boundary additionally enable the controlled conversion between the two kink types.

Taken together, our results show that the deterministic engineering and tailoring of multi-kink states can be achieved by two complementary mechanisms. The $3\mathrm{D}\rightarrow2\mathrm{D}$ planarization of a precursor helix determines the initial defect pattern while helix rotation and $\alpha$ provide control over the number and positions of defects. The subsequent post-selection of kinks and the tuning of the mutual interaction dynamics can then be achieved by using $\alpha$ as a control parameter.
Our findings provide a practical toolbox for reproducible preparation and manipulation of single- and multi-kink states in single-species Coulomb crystals. This enables controlled studies of interacting topological defects, defect-mediated energy transport, and nanofriction.

\section*{Data availability}
The numerical data underlying the results of this study will be made publicly available through the PTB Open Access Repository upon publication~\cite{PTBOAR}. The molecular-dynamics simulations were carried out using custom software developed for laser-cooled ion Coulomb crystals, which is not publicly available. A detailed description of the simulation model, parameters, numerical procedures, and analysis methods is given in the Methods section and Appendices. Additional information regarding the simulation and analysis procedures is available upon reasonable request.

\begin{acknowledgments}
We would like to thank Ramil Nigmatullin for providing the basis of the simulation codes that were used during our research.

We acknowledge support by the projects 18SIB05 ROCIT and 20FUN01 TSCAC.
These projects have received funding from the EMPIR programme cofinanced by the Participating States and from the European Union’s Horizon 2020 research and innovation programme.
Funded by the Deutsche Forschungsgemeinschaft (DFG, German Research Foundation) under Germany’s Excellence Strategy – EXC 2123/2 QuantumFrontiers – 390837967 and through CRC 1227 (DQ-mat), project A07.
\end{acknowledgments}

\bibliography{paper}
\appendix

\section{Electrode geometry and control of helix rotation and planarization quenches}
\label{app:electrode_control}

Following the electrode arrangement of Herschbach et al.~\cite{Herschbach_LinearPaulTrap_2012}, as outlined in Fig.~\ref{fig:trap_schematic}, we consider the effective radial potential near the trap center. The axial confinement along $z$ is generated by neighboring dc segments, is kept constant throughout the protocol, and is not essential for the following discussion. We use the fixed laboratory coordinate system introduced in the main text, where $z$ is the axial trap direction, $x$ and $y$ span the radial plane, and $y$ is aligned with the detection line of sight.

To avoid a static displacement of the crystal, the dc electric field must vanish at the trap center. This is achieved by applying identical voltages to diagonally opposite electrode stacks,
\begin{equation}
U_{\mathrm{t}}^{(1)}=U_{\mathrm{t}}^{(2)}\equiv V_t,\qquad U_{\mathrm{e}}^{(1)}=U_{\mathrm{e}}^{(2)}\equiv V_e.
\label{eq:inv_sym}
\end{equation}
The rf blades carry no additional dc bias. The inversion symmetry imposed by Eq.~\eqref{eq:inv_sym} suppresses dipole fields at the trap center, such that the leading nonconstant dc contribution is quadratic in the radial coordinates.

Near the trap center, the radial potential can therefore be written as
\begin{equation}
V_r(x,y)=\frac{m}{2}
\begin{pmatrix}x&y\end{pmatrix}
\mathbf{K}
\begin{pmatrix}x\\y\end{pmatrix},
\label{eq:radial_potential}
\end{equation}
where
\begin{equation}
\mathbf{K}=
\begin{pmatrix}
K_{xx}&K_{xy}\\
K_{xy}&K_{yy}
\end{pmatrix}
\label{eq:curvature_matrix}
\end{equation}
is the mass-normalized radial curvature matrix expressed in the fixed laboratory basis.

The curvature matrix can be separated into an isotropic contribution and two independent anisotropic components,
\begin{equation}
\mathbf{K}=\overline{K}\,\mathbf{I}+\frac{1}{2}
\begin{pmatrix}
\Delta K&S\\
S&-\Delta K
\end{pmatrix},
\label{eq:curvature_decomposition}
\end{equation}
with
\begin{equation}
\begin{gathered}
\overline{K}=\frac{K_{xx}+K_{yy}}{2},\quad \Delta K=K_{xx}-K_{yy}\\ S=2K_{xy}.
\end{gathered}
\label{eq:curvature_components}
\end{equation}
Here, $\overline{K}$ determines the mean radial confinement, $\Delta K$ describes the laboratory-aligned $x^2-y^2$ anisotropy, and $S$ gives the shear component. The latter two quantities determine the orientation and splitting of the radial principal axes.

Diagonalizing $\mathbf{K}$ gives the orientation $\beta$ of the principal-axis frame relative to the fixed laboratory axes,
\begin{equation}
\tan(2\beta)=\frac{S}{K_{xx}-K_{yy}},
\label{eq:principal_axis_angle}
\end{equation}
and the difference between the two principal curvatures,
\begin{equation}
\delta K=\sqrt{\Delta K^2+S^2}\, .
\label{eq:principal_curvature_splitting}
\end{equation}
Equivalently, the two anisotropic components can be parameterized by
\begin{equation}
\Delta K=\delta K\cos(2\beta),\qquad S=\delta K\sin(2\beta).
\label{eq:anisotropy_angle_parameterization}
\end{equation}
The corresponding principal axes are indicated by $\mathbf{e}_1$ and $\mathbf{e}_2$ in Fig.~\ref{fig:trap_schematic}. The laboratory axes remain fixed and only the principal-axis frame of the radial potential is rotated.

The total curvature can be written as
\begin{equation}
\mathbf{K}=
\mathbf{K}_0
+
\frac{Q}{m} \mathbf{H}_\text{dc}
\label{eq:rf_dc_curvature}
\end{equation}
where $\mathbf{K}_0$ contains all static contributions that remain fixed during the protocol and $\mathbf{H}_\text{dc}$ denotes the Hessian of the static dc potential with its components being
\begin{equation}
H_{ij} \equiv\left.\partial_i\partial_j\Phi_{\mathrm{dc}}\right|_{(0,0)}\,.
\end{equation}

By linear superposition, the static potential generated by the two independently controlled electrode pairs is
\begin{equation}
\Phi_{\mathrm{dc}}(x,y)=V_t\phi_t(x,y)+V_e\phi_e(x,y),
\label{eq:dc_potential_superposition}
\end{equation}
where $\phi_t$ and $\phi_e$ are the unit-voltage potentials generated by the $U_{\mathrm{t}}$ and $U_{\mathrm{e}}$ electrode stacks, respectively, with all other electrodes grounded. Their curvatures are defined as
\begin{equation}
(\phi_\alpha)_{ij}\equiv\left.\partial_i\partial_j\phi_\alpha\right|_{(0,0)},\qquad \alpha\in\{t,e\},
\label{eq:unit_voltage_curvatures}
\end{equation}
such that
\begin{equation}
\Phi_{ij}=V_t(\phi_t)_{ij}+V_e(\phi_e)_{ij}.
\end{equation}

The voltage dependence of the two anisotropic control parameters can then be written as
\begin{equation}
\begin{aligned}
\begin{pmatrix}
\Delta K\\S
\end{pmatrix}
&=
\begin{pmatrix}
\omega_{x}^2-\omega_{y}^2\\0
\end{pmatrix}
\\&+\frac{Q}{m}
\begin{pmatrix}
\Delta\phi_t&\Delta\phi_e\\
2(\phi_t)_{xy}&2(\phi_e)_{xy}
\end{pmatrix}
\begin{pmatrix}
V_t\\V_e
\end{pmatrix},
\end{aligned}
\label{eq:quadrupole_control_matrix}
\end{equation}
where
\begin{equation}
\Delta\phi_\alpha=(\phi_\alpha)_{xx}-(\phi_\alpha)_{yy}.
\end{equation}
Provided that the two electrode responses in Eq.~\eqref{eq:quadrupole_control_matrix} are linearly independent, the voltages $(V_t,V_e)$ can be chosen to generate any desired pair $(\Delta K,S)$ within the experimentally accessible range. The required response matrix can be obtained from an electrostatic field calculation or from an experimental calibration in which the two electrode pairs are varied independently.
\\
\paragraph{Rotation of the helix.}
To control the orientation of the precursor helix, a weak radial anisotropy with principal-curvature splitting $\delta K_{\mathrm{rot}}$ is applied. Rotating the principal-axis frame while keeping this splitting approximately constant requires the two quadrupolar components to be changed jointly according to
\begin{align}
\Delta K_{\mathrm{rot}}(\beta)
&=
\delta K_{\mathrm{rot}}\cos(2\beta)\\
\quad
S_{\mathrm{rot}}(\beta)
&=
\delta K_{\mathrm{rot}}\sin(2\beta).
\label{eq:rotation_components}
\end{align}
The corresponding anisotropic contribution to the curvature matrix is
\begin{equation}
\mathbf{K}_{\mathrm{rot}}(\beta)
=
\frac{\delta K_{\mathrm{rot}}}{2}
\begin{pmatrix}
\cos(2\beta)&\sin(2\beta)\\
\sin(2\beta)&-\cos(2\beta)
\end{pmatrix}.
\label{eq:rotation_curvature_matrix}
\end{equation}
If the principal-axis frame is rotated sufficiently slowly, the helix follows the weak anisotropy and rotates by the same angle $\beta$ about the trap axis. We therefore use $\beta$ for the resulting orientation of the helix relative to the $x$-$z$ detection plane and by that for the controlled preparation of the imprinted kink pattern, as described in Sec.~\ref{subsec:defect_engineering}.
\\
\paragraph{Planarization quench.}
Once the desired helix orientation $\beta$ has been reached, the rotated pinning anisotropy is ramped down while additional confinement is applied along the fixed laboratory $y$ direction. The radial curvature matrix during this handover can be written as
\begin{equation}
\mathbf{K}(t)=K_0(t)\mathbf{I}+a(t)\mathbf{K}_{\mathrm{rot}}(\beta)+q(t)
\begin{pmatrix}
0&0\\
0&1
\end{pmatrix}.
\label{eq:combined_rotation_quench}
\end{equation}
Here, $q(t)$ directly represents the additional curvature along the fixed $y$ direction. Since $\mathbf{K}_{\mathrm{rot}}$ is traceless, the mean radial curvature defined in Eq.~\eqref{eq:curvature_components} is
\begin{equation}
\overline{K}(t)=K_0(t)+\frac{q(t)}{2}.
\label{eq:mean_curvature_handover}
\end{equation}
The corresponding anisotropic components are
\begin{equation}
\Delta K(t)=a(t)\delta K_{\mathrm{rot}}\cos(2\beta)-q(t),\qquad S(t)=a(t)\delta K_{\mathrm{rot}}\sin(2\beta).
\label{eq:combined_anisotropic_components}
\end{equation}
The functions $a(t)$ and $q(t)$ specify the handover protocol. Their endpoint values are fixed by
\begin{equation}
\begin{gathered}
a(0)=1,\quad a(t_Q)=0,\\q(0)=0,\quad q(t_Q)=q_f,
\end{gathered}
\label{eq:handover_boundary_conditions}
\end{equation}
where their time dependence can be chosen according to the desired adiabatic or finite-rate planarization regime discussed in Dec.~\ref{subsec:defect_engineering}. In an experimental implementation, the target values $\Delta K(t)$ and $S(t)$ from Eq.~\eqref{eq:combined_anisotropic_components} are converted at each time into the corresponding electrode voltages $V_t(t)$ and $V_e(t)$ through the relation in Eq.~\eqref{eq:quadrupole_control_matrix}.

At the end of the protocol, the principal axes then coincide with the fixed laboratory axes and the crystal is confined to the observable $x$-$z$ plane. In the idealized zero-temperature limit, the two ramps could be applied sequentially. Experimentally, however, they should be overlapped and performed without an extended interval of near-radial symmetry, since thermal fluctuations may otherwise induce an uncontrolled global rotation of the helix.

\section{Full dependence and phenomenological fits of stability boundaries}
\label{app:stability_fits}

In Sec.~\ref{sec:post_selection} we use defect-specific stability regions as a post-selection criterion. Here we provide the full dependence $z_{\max}(N,\alpha)$ with phenomenological parametrizations of the extracted stability boundaries. The full dependence for both odd and extended kinks is extracted from numerically calculated PN potentials for different ion numbers $N$ and trap anisotropies $\alpha$.

\subsection{Odd kinks}
\begin{figure}
  \includegraphics[width=\columnwidth]{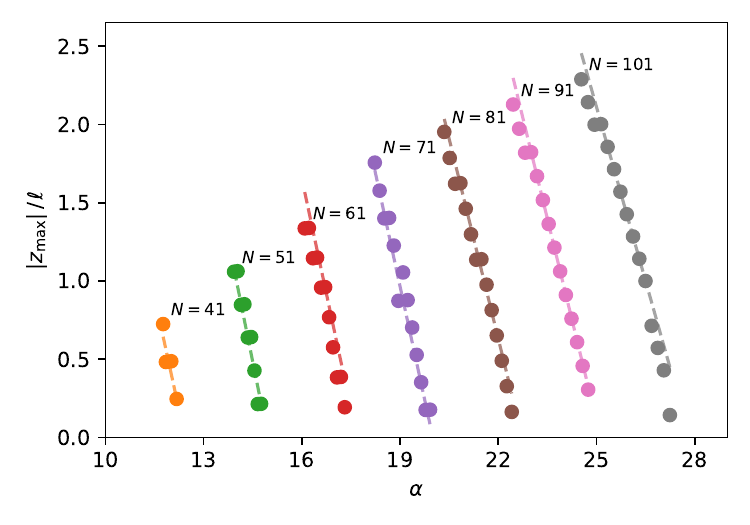}
  \caption{\justifying Maximum axial position $z_\text{max}$ at which an odd kink can be trapped at $T=0$ as a function of the particle number $N$ and the ratio of trapping frequencies $\alpha$. The dashed lines show linear fits to emphasize the linear trend in the data for each crystal size.
}
  \label{fig:odd_kink_range}
\end{figure}
\begin{figure}
  \includegraphics[width=\columnwidth]{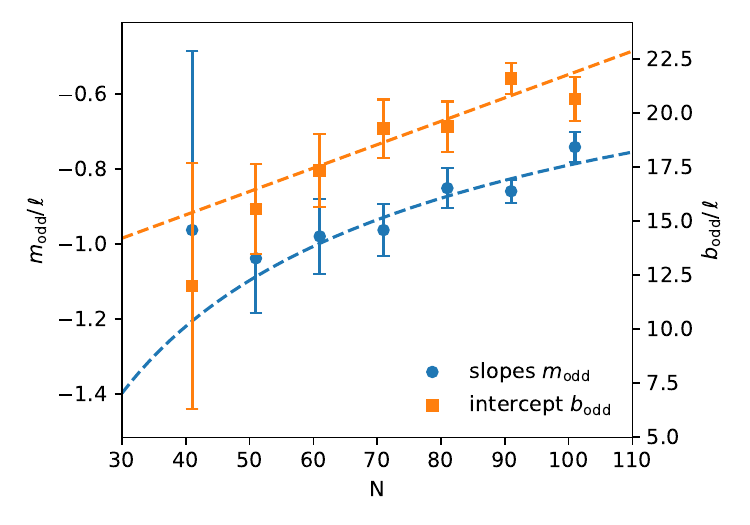}
  \caption{\justifying
  Fit parameters of the linear trends used in Fig.~\ref{fig:odd_kink_range}, namely the slope $m_\text{odd}$ and intercept $b_\text{odd}$, shown as a function of ion number $N$. These trends provide a compact summary of the dependence of the odd-kink stability boundary $z_\text{max}(N,\alpha)$ on $\alpha$ and $N$ within the investigated parameter range. The explicit fit functions are given in the text.}
  \label{fig:odd_trend_fits}
\end{figure}
The extracted stability boundary for odd kinks is shown in
Fig.~\ref{fig:odd_kink_range}. Within the investigated parameter
range, the numerically extracted boundary at fixed $N$ is well
approximated by a linear dependence on $\alpha$,
\begin{equation}
    z_{\max}(N,\alpha) \, / \, \ell
    \approx
    m_{\mathrm{odd}}(N)\alpha
    + b_{\mathrm{odd}}(N),
    \label{eq:zmax_envelope}
\end{equation}
with the $N$-dependent slope and intercept parametrized as
\begin{align}
    m_{\mathrm{odd}}(N)
    &= -A_{\mathrm{odd}} N^{-p_{\mathrm{odd}}},
    \label{eq:modd_of_N}
    \\[4pt]
    b_{\mathrm{odd}}(N)
    &= s_{\mathrm{odd}}N+b_{\mathrm{odd},0}.
    \label{eq:bodd_of_N}
\end{align}
The fitted parameters are
\begin{align*}
    A_{\mathrm{odd}} &= 7.0 \pm 4.4, &p_{\mathrm{odd}} &= 0.47 \pm 0.14,\\
    s_{\mathrm{odd}} &= 0.108 \pm 0.032, &b_{\mathrm{odd},0} &= 10.9 \pm 2.8.
\end{align*}
The uncertainties denote one standard deviation obtained from the
fit covariance matrices. These parametrizations are illustrated in
Fig.~\ref{fig:odd_trend_fits}.
Because the PN landscape is periodic on the lattice scale, stable trapping sites occur only at discrete axial positions separated by approximately the local inter-ion spacing along $z$. As a
result, the extracted boundary exhibits a staircase-like structure when $\alpha$ is varied.

\subsection{Extended kinks}
\begin{figure*}
  \includegraphics[width=0.85\textwidth]{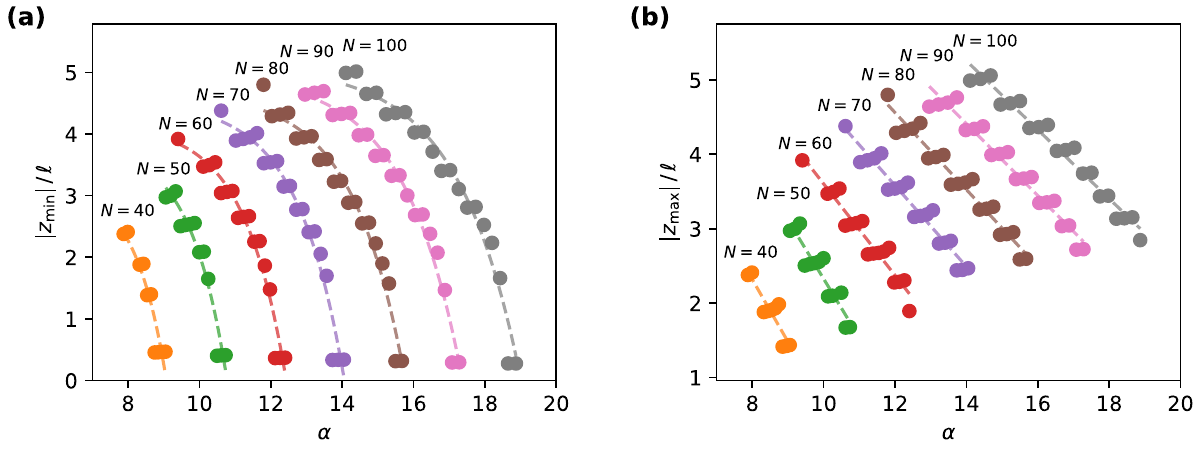}
  \caption{\justifying\textbf{(a)} Smallest axial position (excluding the global minimum at $z=0$) at which an extended kink can be trapped based on the energy threshold given in \eqref{eq:trapping_threshold}, shown for different system sizes and ratios of the trapping frequencies $\alpha$. \textbf{(b)} Maximum axial position $z_\text{max}$ at which an extended kink can be trapped as a function of the particle number $N$ and the ratio of trapping frequencies $\alpha$. All data were obtained by calculating the Peierls–-Nabarro potential for a single extended kink for the given crystal size. The $\alpha$ values were chosen evenly spaced within the extended regime (see Fig.~\ref{fig:alpha_vs_Nions}, neglecting values close to the transition regions. Dashed lines show exponential fits to emphasize the trend in the data for each crystal size.
}
  \label{fig:ext_kink_range}
\end{figure*}

The stability boundaries for the extended case are shown in Fig.~\ref{fig:ext_kink_range}. The outer boundary $z_\text{max}$ is defined analogously to the odd-kink case as the largest axial position at which a local PN minimum fulfilling the threshold condition exists. The inner boundary $z_\text{min}$ is defined as the stable trapping minimum closest to the global PN minimum at $z=0$. 

\begin{figure*}
  \includegraphics[width=0.9\textwidth]{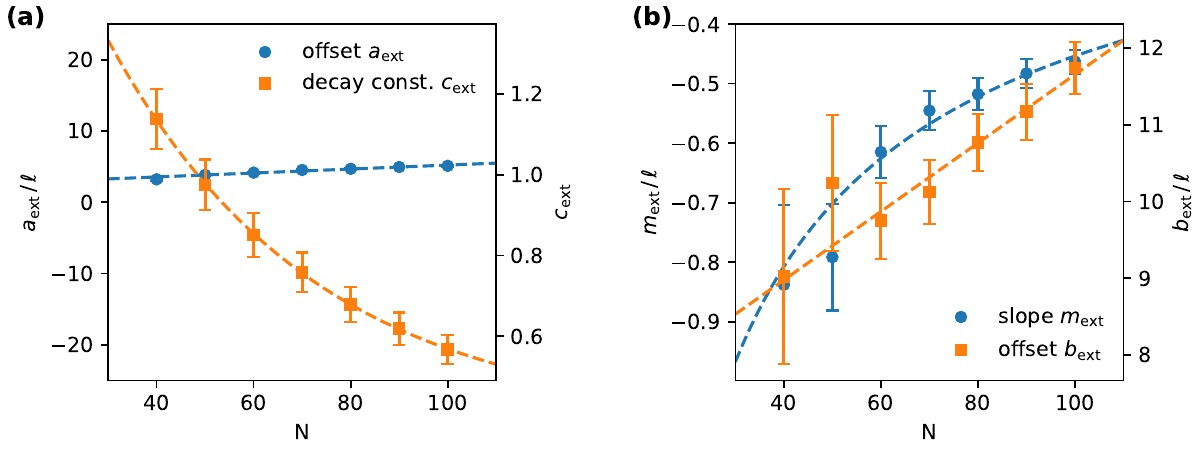}
  \caption{\justifying
  Fit parameters summarizing the stability bounds of extended kinks extracted from Fig.~\ref{fig:ext_kink_range}.
  \textbf{(a)} Parameters $a_\text{ext}(N)$ and $c_\text{ext}(N)$ entering the exponential trend used for the inner boundary $z_\text{min}$ (cf.~Fig.~\ref{fig:ext_kink_range}(a)) together with the phenomenological fits described in the text.
  \textbf{(b)} Parameters $m_\text{ext}(N)$ and $b_\text{ext}(N)$ entering the linear trend used for the outer boundary $z_\text{max}$ (cf.~Fig.~\ref{fig:ext_kink_range}(b)) together with the phenomenological fits described in the text.}
  \label{fig:ext_trend_fits}
\end{figure*}

Extended kinks exhibit a stability region bounded by an inner and an
outer edge,
\begin{equation}
    z_{\min}(N,\alpha)
    \leq |z|
    \leq z_{\max}(N,\alpha),
\end{equation}
as shown in Fig.~\ref{fig:ext_kink_range}. In the explored parameter
range, the shift of the inner boundary with increasing $\alpha$ is
captured by the exponential trend
\begin{equation}
    |z_{\min}|(N,\alpha) \, / \, \ell
    \approx
    a_{\mathrm{ext}}(N)
    - A_{\min}\exp\!\left[c_{\mathrm{ext}}(N)\alpha\right],
    \label{eq:zmin_exp}
\end{equation}
where $A_{\min}$ is a common prefactor. The $N$-dependent fit
parameters are parametrized as
\begin{align}
    a_{\mathrm{ext}}(N)
    &= s_a N + a_0,
    \label{eq:aext_of_N}
    \\[4pt]
    c_{\mathrm{ext}}(N)
    &= c_{\infty}
    + c_0\exp\!\left(-\lambda N\right).
    \label{eq:cext_of_N}
\end{align}
The fitted parameters are
\begin{align*}
    A_{\min} &= (1.1 \pm 0.7)\times 10^{-4}, &\; s_a &= 0.0278 \pm 0.0030,\\
    a_0 &= 2.44 \pm 0.23, &\; c_{\infty} &= 0.386 \pm 0.202,\\c_0 &= 1.916 \pm 0.626, &\; \lambda  &= 0.0235 \pm 0.0128.
\end{align*}
These parametrizations are illustrated in
Fig.~\ref{fig:ext_trend_fits}(a).

The outer boundary $z_{\max}(N,\alpha)$ varies more weakly with
$\alpha$ and is, over the investigated range, well described by the
approximately linear trend
\begin{equation}
    |z_{\max}|(N,\alpha) \, / \, \ell
    \approx
    m_{\mathrm{ext}}(N)\alpha
    + b_{\mathrm{ext}}(N).
    \label{eq:zmax_ext}
\end{equation}
The $N$-dependent slope and intercept are parametrized as
\begin{align}
    m_{\mathrm{ext}}(N)
    &= -A_{\max}N^{-p_{\mathrm{ext}}},
    \label{eq:mext_of_N}
    \\[4pt]
    b_{\mathrm{ext}}(N)
    &= s_b N+b_0.
    \label{eq:bext_of_N}
\end{align}
The fitted parameters are
\begin{align}
    A_{\max}
    &= 8.23 \pm 3.95,
    &
    p_{\mathrm{ext}}
    &= 0.629 \pm 0.109,
    \\
    s_b
    &= 0.0446 \pm 0.0107,
    &
    b_0
    &= 7.19 \pm 0.88.
\end{align}
These parametrizations are illustrated in
Fig.~\ref{fig:ext_trend_fits}(b). The quoted uncertainties denote one
standard deviation obtained from the fit covariance matrices.

The outer stability boundary $z_\text{max}$ exhibits a markedly weaker dependence on $\alpha$ and, within the explored parameter range, follows an approximately linear trend similar to that observed for odd kinks.

Together, the quantities $z_\text{min}$ and $z_\text{max}$ define the axial region in which stable trapping of extended kinks is possible. As in the odd-kink case, the admissible trapping positions occur only at discrete axial locations set by the underlying lattice structure. For extended kinks, these positions are separated by approximately twice the local inter-ion  spacing, reflecting the doubled periodicity of the PN potential.

\section{Two-kink configurations and definition of kink centers}
\label{app:multi_kink_position}
For odd kinks, the defects are sharply localized and their positions can be defined unambiguously from the sign change of the transverse zigzag orientation.

For extended kinks, the distortion typically spans several lattice sites. In order to accurately determine the positions of multiple extended kinks in our crystals, we construct the axial bond–distortion profile
\begin{equation}
e_j = \left[(z_{j+1}-z_j) - (z^{(0)}_{j+1}-z^{(0)}_j)\right]^2,
\end{equation}
where $z_j$ are the axial coordinates and $z^{(0)}_j$ denote the no-kink reference configuration. 
The two largest local maxima of $e_j$, separated by a minimal bond distance, are used as anchors for the left and right extended kinks. 
Around each anchor position $w$, we apply a Gaussian window $W(w)$ and define continuous kink centers via weighted centroids,
\begin{equation}
z_k =
\frac{\sum_j w_j\, e_j\, W(w_j)}
     {\sum_j e_j\, W(w_j)}.
\end{equation}
In contrast to the global collective coordinate of Ref.~\cite{PartnerDefects2013a}, which yields a single defect coordinate and can become ambiguous in the presence of multiple defects, our windowed definition provides two localized position coordinates and remains well-defined as long as the two distortion peaks are resolvable.

\section{Defect regimes}
\label{app:defect_regimes}
\begin{figure}
  \includegraphics[width=\columnwidth]{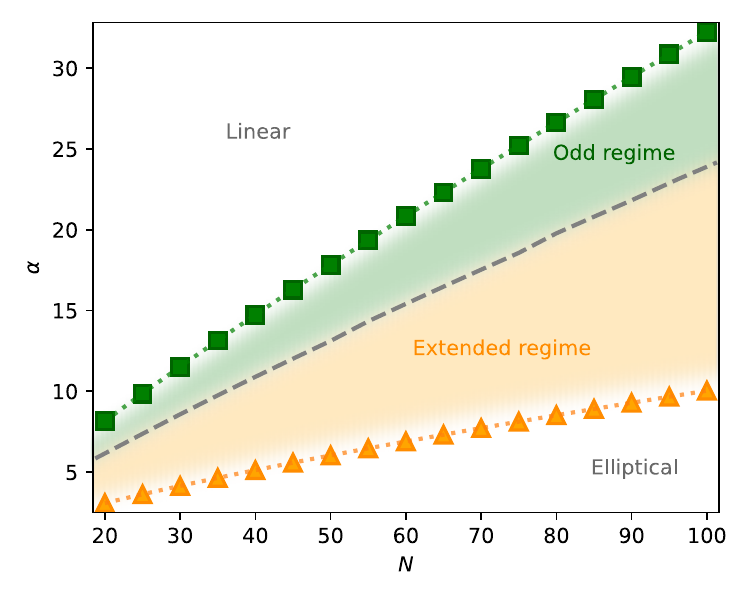}
  \caption{\justifying Topological defect regimes as a function of the trapping anisotropy $\alpha$ and the number of ions $N$, evaluated at the center of the crystal. The central gray dashed line marks the transition boundary between the extended and odd regime, where the ratio of the transverse separation $a$ and the axial spacing $b$ is $a/b=1$. Defects close to this transition boundary are in the intermediate regime where their dynamics are characterized by both transverse and axial motion of the ions~\cite{PartnerDefects2013a}. The green squares mark the linear to zigzag transition, while orange triangles mark the transition from the zigzag phase into a 2D elliptical phase~\cite{Yan_StructuralPhases_2016}.}
  \label{fig:alpha_vs_Nions}
\end{figure}
The defects can be broadly classified into two types: odd and extended kinks. The emergence and stability of these defects depend on the anisotropy $\alpha$ and the number of ions $N$ in the crystal. Both odd and extended kinks can occur in the 2D zigzag phase but they occupy different parameter regimes ($N$, $\alpha$) within that phase.
The odd and extended regimes can be classified by the ratio $a/b$, where $a$ is the transverse separation and $b$ the axial spacing between the ions. The odd regime is characterized by $a/b < 1$ and for extended kinks by $a/b>1$~\cite{PartnerDefects2013a}. Due to the inhomogeneous axial spacing, in a finite crystal, both regimes can coexist at different axial positions. Fig.~\ref{fig:alpha_vs_Nions} illustrates both regimes in the planar case ($ \omega_z \ll \omega_x < \omega_y $) with $a$ and $b$ being evaluated at the center of the crystal $z=0$ \cite{PartnerDefects2013a, Yan_StructuralPhases_2016}. The propagation of a kink through the zigzag lattice in the odd regime is characterized by significant transverse motion of the ions close to the kink center. In the extended regime, the defect center propagates by ions moving primarily along the axial direction~\cite{PartnerDefects2013a}. 
\end{document}